# Ephemeris Refinement for Qatar-4 b, HAT-P-18 b, and CoRoT-1 b with Small Telescope and TESS Observations

**Heather B. Hewitt**
**Mikaela Chamlee**
**Trinity Hockman**
**Annalise Igyarto**
**Dan van der Burg**
**Federico R. Noguer**
**Ira Bell**
**Jeremie Abides**
**Simon Abshear**
**Philip Arthur Ellis**
**Shashank Avinash Araokar**
**Eurico Bras**
**Jamie Buckingham**
**Dion Bumpus**
**Kartavya Chauhan**
**Logan Conrad**
**Koty Creel**
**Courtney Dammann**
**Samantha Duarte Rodriguez**
**Jeffrey Ensor**
**Benjamen Erickson**
**Adam Facciponti**
**Logan Farrar**
**Blaise Foca**
**Holly Foreman**
**Mark Foreman**
**Haley Fournier**
**Jeff Franklin**
**Keith Goss**
**Dustin Hubbartt**
**Nicolette Jeantilus**
**Justin Kobrin**
**Spencer Lunsford**
**Wyatt McCormick**
**Zachary McNeely**
**Kurtis Merrill**
**Sarah Miller**
**Dean Miller**
**Sofie Miura**
**Cody Montgomery**
**Karina Munoz**
**Keith Norton**
**Anna Plum**
**Jake Raby**
**Jay Thompson**
**Sohley Wallace-Pichardo**
**Molly N. Simon**
*School of Earth and Space Exploration, Arizona State University, 781 E Terrace Mall, Tempe, AZ 85287-6004; hbhewitt@asu.edu*

**Robert T. Zellem**
*NASA Goddard Space Flight Center, Exoplanets and Stellar Astrophysics Laboratory (Code 667), Greenbelt, MD 20771*

**Abstract** We present updated transit timing measurements for three hot Jupiters (Qatar-4 b, HAT-P-18 b, and CoRoT-1 b) by leveraging data collected from the MicroObservatory Telescope Network, a network of small, robotic ground-based telescopes, and the NASA Transiting Exoplanet Survey Satellite (TESS). By combining these data with archival published results, we present the most precise orbital solutions to date for all three systems, allowing for precise transit time predictions for future missions. We report an updated mid-transit time for Qatar-4 b of $2458919.5838 \pm 0.000089$ $BJD_{TDB}$ and an updated orbital period of $1.80536560 \pm 0.00000021$ days. For HAT-P-18 b, we find a mid-transit time of $2459743.85340 \pm 0.000022$ $BJD_{TDB}$ and an updated orbital period of $5.50802957 \pm 0.00000012$ days. For CoRoT-1 b, we report a mid-transit time of $2456268.99083 \pm 0.000099$ $BJD_{TDB}$ and an updated orbital period of $1.50896846 \pm 0.000000071$ days. Our results demonstrate improvements over recently published ephemerides, with reductions of 36.4%, 4.35%, and 17.5% in mid-transit time uncertainties and 65.0%, 77.4%, and 16.9% in orbital period uncertainties for Qatar-4 b, HAT-P-18 b, and CoRoT-1 b, respectively. The results of this study improve the precision of future transit predictions and demonstrate the value of coordinated small-telescope monitoring (and citizen science initiatives) when updating the orbital parameters of hot Jupiters.

## 1. Introduction

The rapid pace of exoplanet discovery has offered both unprecedented opportunities and logistical challenges for the astronomical community. There have been over 6,000 confirmed exoplanets discovered with over 7,000 candidates still to be validated. As time progresses, ephemeris predictions degrade as observational uncertainties compound with each orbit, leading to increasingly imprecise transit time predictions (Fowler *et al.* 2021; Zellem *et al.* 2020). Hot Jupiters, while being one of the most extensively studied classes of exoplanets, are also particularly vulnerable to their mid-transit time uncertainties degrading more rapidly than long-period planets due to their proximity to their host star(s) (Zellem *et al.* 2020). Consequently, regular follow-up observations are necessary to maintain precise mid-transit times. This effort is crucial in enabling precise future transit time predictions for subsequent studies using telescopes such as the Hubble Space Telescope (HST), the James Webb Space Telescope (JWST), or the upcoming Atmospheric Remote-sensing Infrared Exoplanet Large-survey (Ariel) mission. The efficient scheduling of these large space-based telescopes is necessary to limit wasted time and maximize science returns.

The sheer volume of targets requiring follow-up observations make using solely space- or large, ground-based telescopes impractical due to their strict temporal and financial constraints. For example, thousands of planet candidates and confirmed exoplanets have been identified by TESS that will require follow-up observations to maintain accurate transit ephemerides (Guerrero *et al.* 2021). Leveraging data collected by citizen science initiatives such as Exoplanet Watch (NASA 2026) can address the challenges of accessing large ground- and space-based telescope observations to maintain mid-transit times. Exoplanet Watch engages citizen scientists in collecting and analyzing data with small, ground-based telescopes such as the 6-inch robotic telescopes from the MicroObservatory Robotic Telescope Network (Sadler *et al.* 2001). The approach of integrating large volumes of citizen science observations with publicly available space-based telescope observations and published results from previous studies using high-precision instruments help to build a broader observational baseline (e.g., Noguer *et al.* 2024). Studies have shown that using these expanded data sets improves mid-transit time precision compared to using smaller datasets from high-precision telescopes alone (Hewitt *et al.* 2023a, 2024; Mizrachi *et al.* 2021; Pearson *et al.* 2022; Torres *et al.* 2025). Additionally, the NASA Transiting Exoplanet Survey Satellite (TESS; Ricker *et al.* 2015) continuously generates high-quality light curves for known exoplanets during its all-sky survey, including systems that present observational challenges from Earth's surface. Leveraging these space-based observations in combination with more readily accessible ground-based data allows for routine ephemeris refinement of many hot Jupiters (e.g., Ivshina and Winn 2022; Kokori *et al.* 2023; Noguer *et al.* 2024; Sgro *et al.* 2024).

In this study, we refined the mid-transit times and orbital periods of Qatar-4 b, HAT-P-18 b, and CoRoT-1 b by combining small, ground-based robotic telescope observations with new TESS data and previously published results, demonstrating that coordinated citizen-science monitoring can meaningfully improve ephemerides beyond studies relying primarily on space-based or professional observatory data alone. The work done in this study represents a culmination of the efforts completed across three semesters of an online Course-based Undergraduate Research Experience (CURE) as part of the online Astronomy and Planetary Science degree program at Arizona State University. The course was developed as a way to give online students the opportunity to engage in authentic research. Details of the course structure and impacts on student affective outcomes are discussed in Hewitt *et al.* (2023b).

## 2. Ground-based observatories

Our data were primarily acquired using the MicroObservatory Robotic Telescope Network developed and operated by the Harvard Smithsonian Center for Astrophysics. The MicroObservatory Robotic Telescope Network is designed to allow students, educators, and other users to capture and analyze images of celestial objects from anywhere with an internet connection, without needing their own telescope (Sadler *et al.* 2001). These telescopes are positioned at observatories linked with the Harvard College Observatory in Massachusetts and the Whipple Observatory in Arizona. The observations used in this study were collected with the "*Cecilia*" and "*Ben*" telescopes located at the Fred Lawrence Whipple Observatory on Mt. Hopkins, at an elevation of approximately 2590 meters.

Like all the MicroObservatory telescopes, *Cecilia* and *Ben* are identical, entirely self-contained systems designed to prevent weather-related damage and ensure nightly observations for extended periods with minimal maintenance. *Cecilia* and *Ben* are both custom-designed Maksutov-Newtonian telescopes, featuring a primary mirror with a diameter of 6 inches and a focal length of 560 mm. The telescope is paired with a specialized Kodak KAF1400 CCD imaging sensor, which has pixels measuring 6.9 microns on each side, resulting in an imaging field of view measuring $0.94 \times 0.72$ degrees. A $2 \times 2$ pixel binning technique was applied, producing an effective resolution of 5.0 arc-seconds per pixel (Sadler *et al.* 2001). Imaging was performed utilizing a clear (visible) filter with exposure durations of 60 seconds.

Five additional CoRoT-1 b observations were obtained using a privately operated Planewave CDK24 telescope hosted at Sierra Remote Observatories (California, USA). The 24-inch (0.61 m) f/6.5 system was paired with an SBIG Aluma AC4040 CMOS camera ($4096 \times 4096$ pixels; 9 μm pixel size; typical read noise 4.5 $e^-$) and operated with thermoelectric cooling. Exposure times ranged from 0.01 to 3600 s.

## 3. Data

### 3.1. Ground-based data

For the photometric evaluation of all ground-based data in this study, including observations from both MicroObservatory and the private observatory, we used the EXOplanet Transit Interpretation Code (EXOTIC; Zellem *et al.* 2020), a Python 3 pipeline developed and maintained by Exoplanet Watch. EXOTIC processes raw photometric images in FITS format by tracking the positions of target and comparison stars across all frames and performing aperture photometry to produce time-series light curves. Model parameters are estimated using a Bayesian fitting routine. We allowed the following parameters to vary in the fitting process: mid-transit time, planet-to-star radius ratio, orbital inclination, transit duration, and airmass correction terms. A full description of the EXOTIC fitting procedure is outlined in Zellem *et al.* (2020). This software can be operated both on local machines and cloud platforms, such as Google Colaboratory (Colab). Our team opted to use the Google Colab platform to run EXOTIC, leveraging its collaborative features for ease of sharing data and data products between researchers.

We employed a two-stage quality control process following the standardized methodology described in Hewitt *et al.* (2024, 2023a). First, we performed an initial reduction of all observations to identify potentially significant ($> 3\sigma$) light curves. We defined statistical significance by the ratio of the measured transit depth to its associated uncertainty. All images were then re-examined by subsets of the authors to ensure consistent quality standards across all reductions. Images were evaluated based on several criteria: cloud coverage, visual noise, atmospheric haze, streaking from tracking errors, and adverse weather conditions. Images were excluded from the final dataset if any of these factors compromised our ability to reliably identify the target star or surrounding comparison stars.

For each target, we identified suitable comparison stars for EXOTIC's photometric fitting procedure. All comparison stars for Qatar-4 b and HAT-P-18 b were selected from the American Association of Variable Star Observers (AAVSO) recommendations and verified to be non-variable. For CoRoT-1 b, with no AAVSO-recommended comparison stars available in the field of view, we selected comparison stars located near CoRoT-1 b to ensure they were affected similarly by systematic effects, such as variations in air mass. Figure 1 presents the comparison star positions within the field of view for each target as generated by the AAVSO Variable Star Plotter (VSP; Craig *et al.* 2023).

In this study, we reduced 14 statistically significant light curves of CoRoT-1 b obtained using data from the MicroObservatory Robotic Telescope Network and the Planewave CDK24 telescope. We also reduced 17 statistically significant light curves of Qatar-4 b and 10 of HAT-P-18 b, all obtained using the MicroObservatory. Representative light curves from MicroObservatory observations for each target are shown in Figure 2, while the full set of light curves reduced using MicroObservatory and the Planewave CDK24 is presented in Appendix A. A summary of the mid-transit times and transit depths for all significant ground-based observations is presented in Table 1.

3.2. Transiting Exoplanet Survey Satellite (TESS) data

Qatar-4 b, HAT-P-18 b, and CoRoT-1 b were observed by TESS in Sectors 84 (2024 October 1–October 26), 79 (2024 May 21–June 02), and 87 (2024 December 18–2025 January 14), respectively, at 2-minute cadences. These observations occurred after the most recently published ephemeris updates that utilized TESS data for each system (Ivshina and Winn 2022; Kokori *et al.* 2023). The calibrated time-series data analyzed in this work were retrieved from the Mikulski Archive for Space Telescopes (MAST; STScI 2024) and processed following the light-curve extraction and cleaning procedure described by Noguer *et al.* (2024). This method incorporates an updated aperture selection scheme based on Pearson (2019), which mitigates pointing-induced systematics. Data obtained within 30 minutes of TESS momentum dump events were excluded due to elevated systematic noise. Additionally, we removed 3σ outliers using a rolling median filter with a 32-minute window. Quality-flagged data points were also excluded. Long-term stellar variability was detrended using the PYTHON package WOTAN (Hippke *et al.* 2019), with detrending parameters chosen to preserve the planetary transit signals. Example light curves for each target are presented in Figure 3 with the full set of TESS data presented in Appendix B. The resulting mid-transit times and transit depths are presented in Table 2.

## 4. Analysis

In order to derive updated mid-transit times, we created Observed – Calculated (O–C) plots for each of our targets. In addition to the mid-transit times reduced in this study (Tables 1 and 2), we also included those derived in previous studies. For Qatar-4 b, we included mid-transit times from Kokori *et al.* (2022, 2023) and Ivshina and Winn (2022) fits of Wang *et al.* (2021), Mallonn *et al.* (2019), and Alsubai *et al.* (2017). The O–C plot for HAT-P-18 b included published mid-transit times

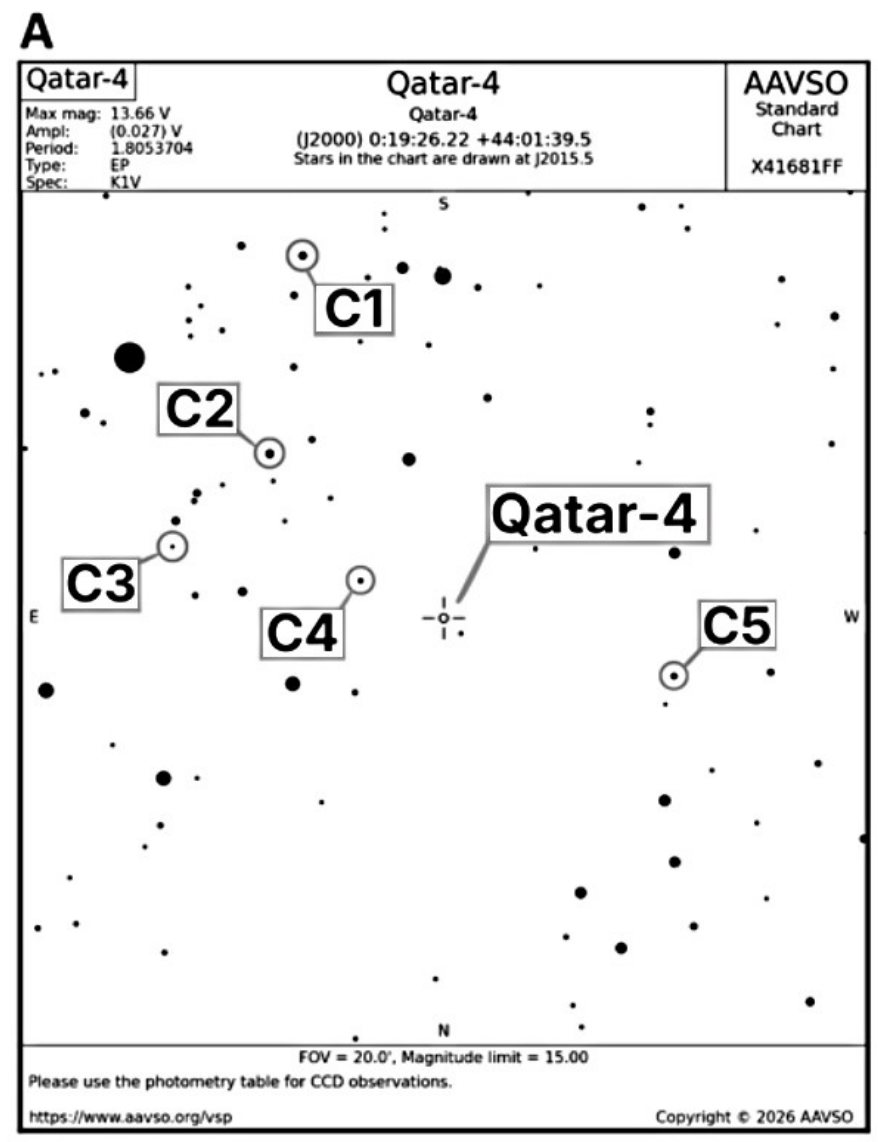

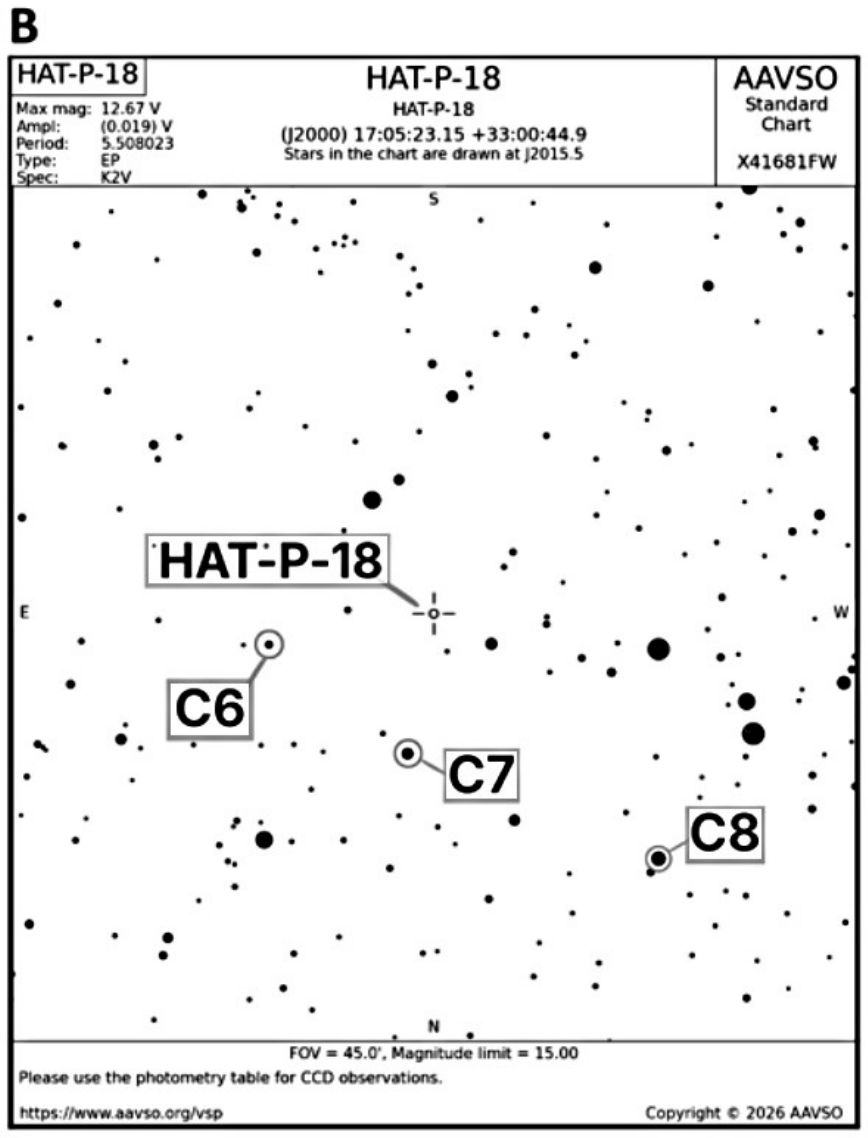

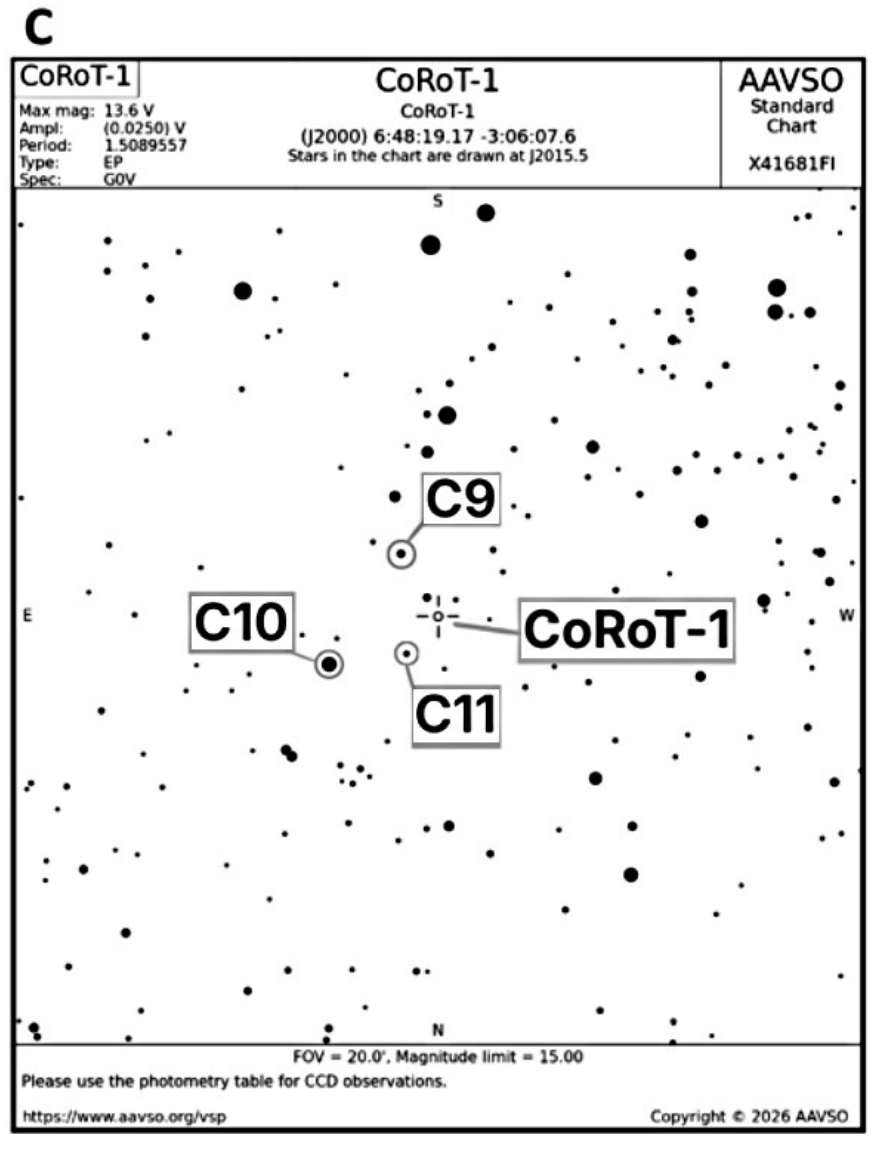

Figure 1. AAVSO VSP star charts for Qatar-4 b (A), HAT-P-18 b (B), and CoRoT-1 b (C). The field of views are 20', 45', and 20', respectively. The comparison stars used in this study are labeled C-1 through C-11.

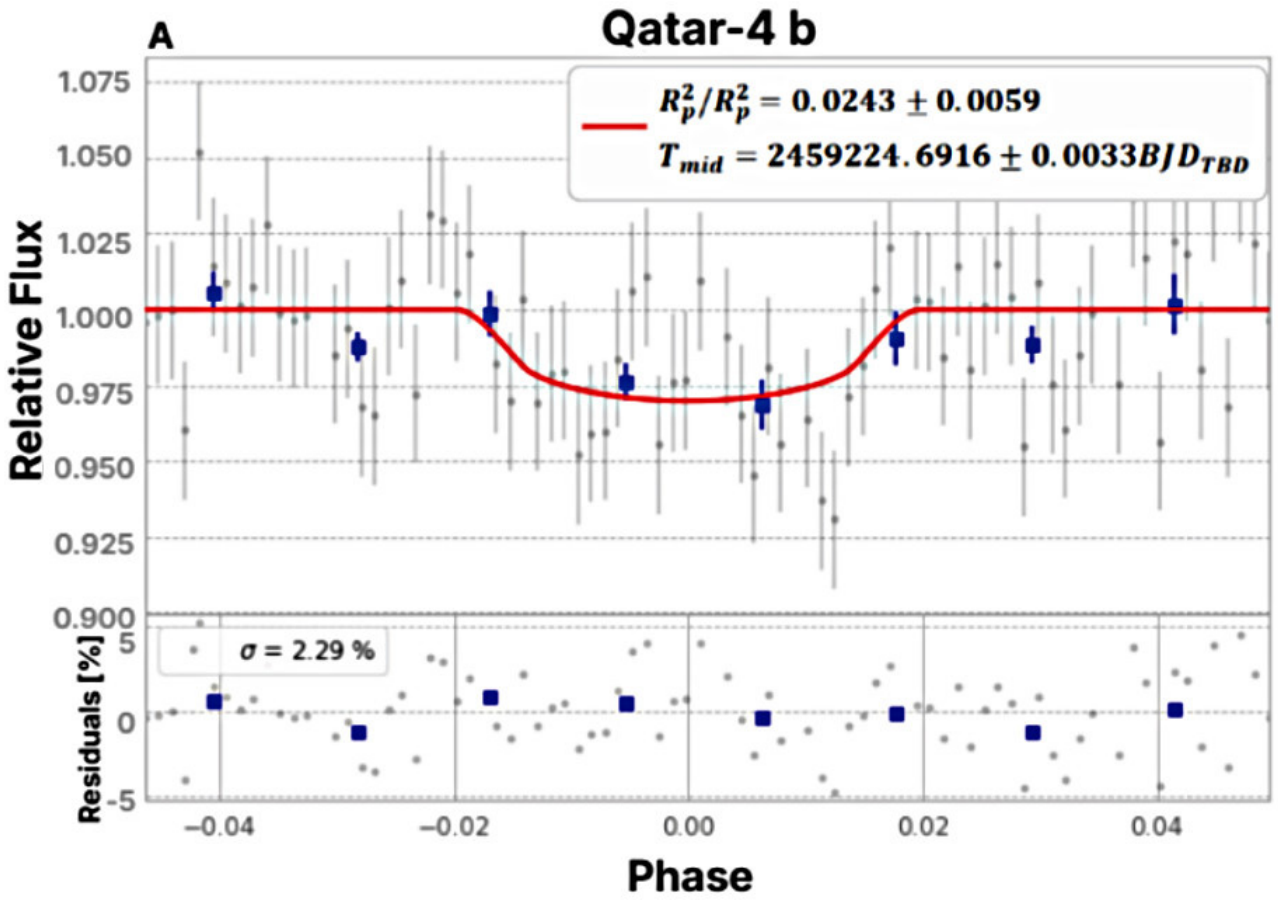

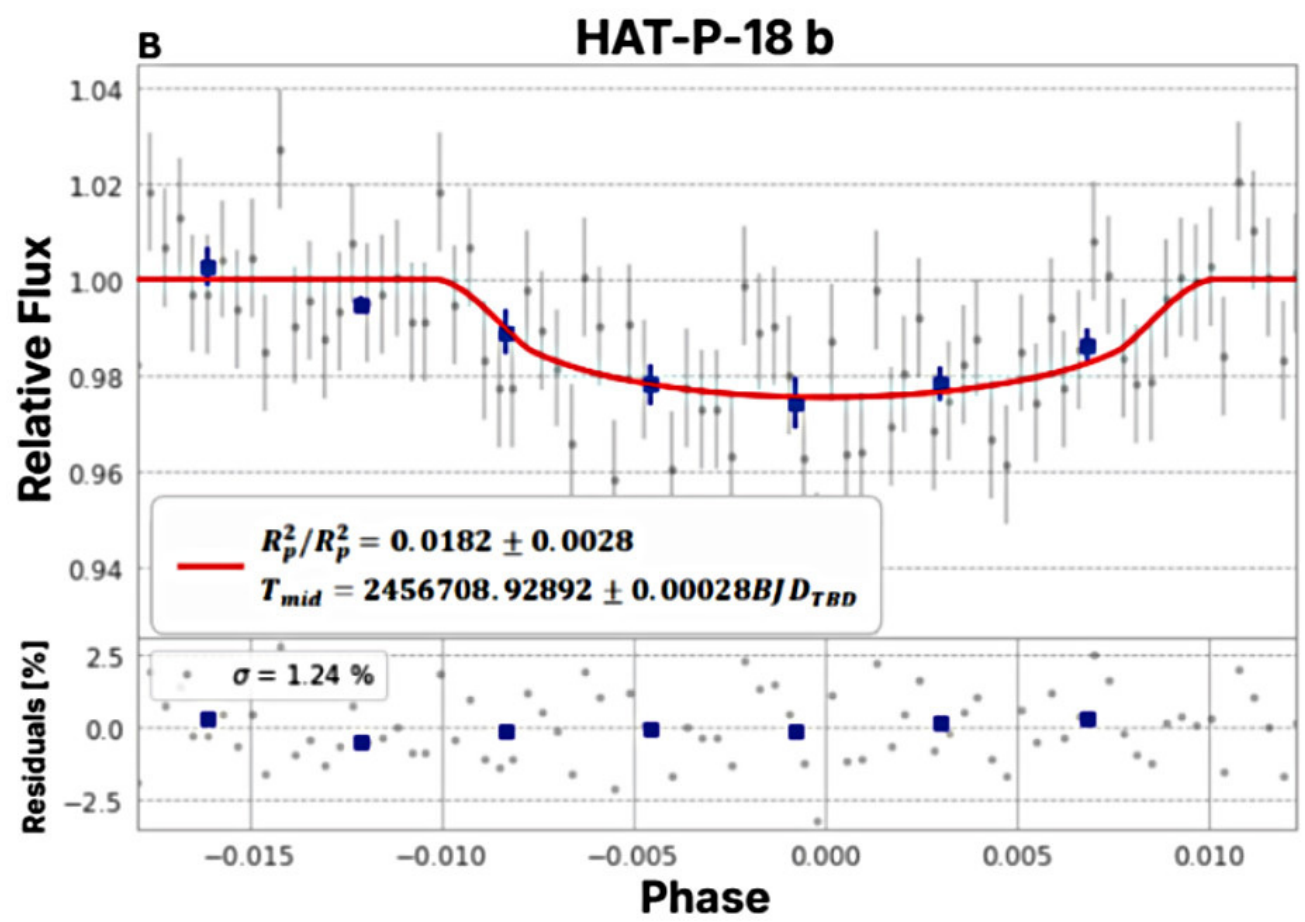

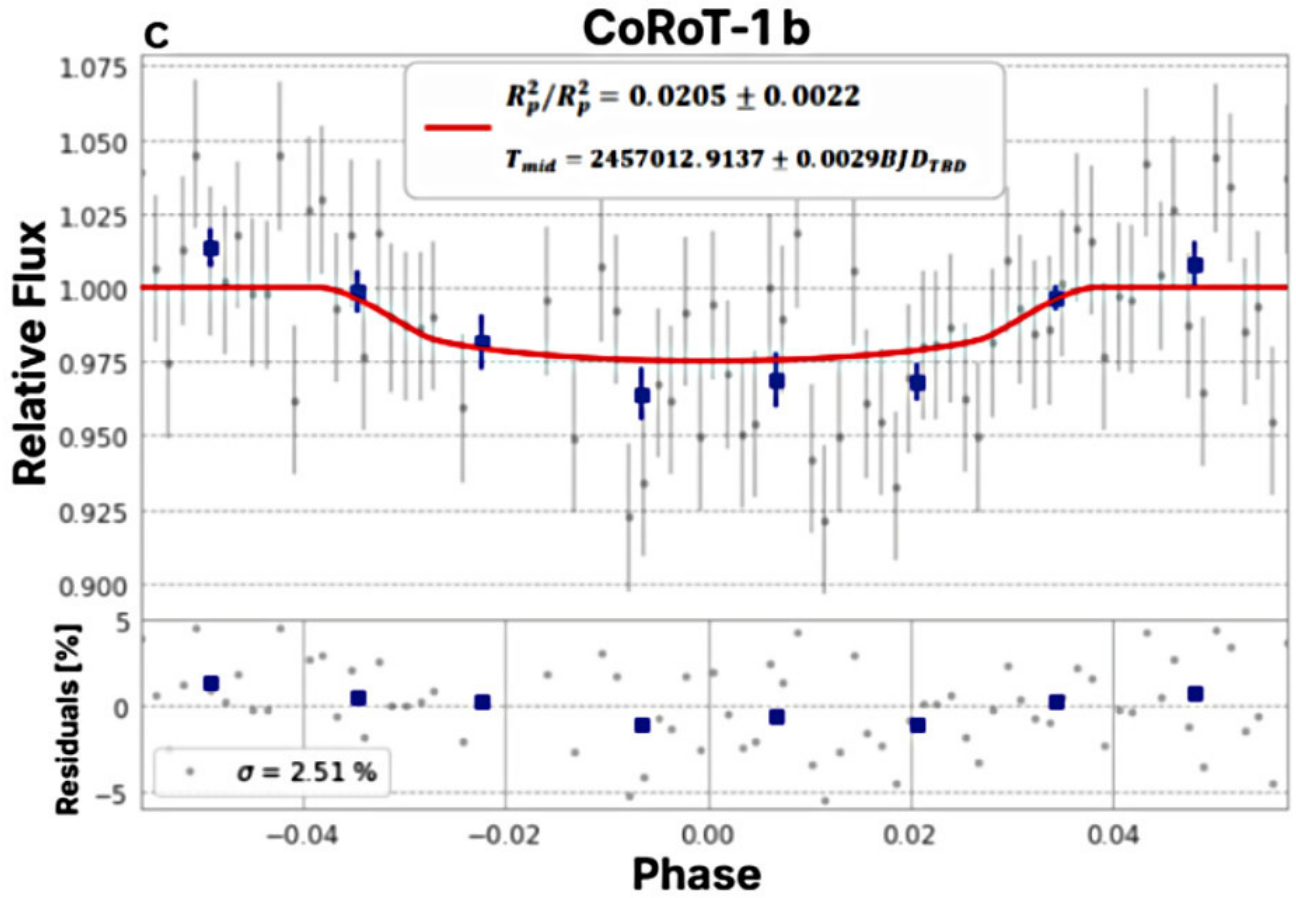

Figure 2. Example light curves reduced using MicroObservatory data for each target. Qatar-4 b (A) was observed on 2021-01-10, HAT-P-18 b (B) was observed on 2014-02-09, and CoRoT-1 b (C) was observed on 2014-12-20. The gray data points represent data collected from each image of the data set. The blue data points represent an average from a set of binned data points and were used to fit the light curve.

Table 1. Mid-transit times and transit depths for the statistically significant detections (> 3σ) reduced in this study using ground-based data.

| *Target* | *Date* | *Mid-Transit Time ($BJD_{TDB}$)* | *$Rp^2/R_s^2$* |
|---|---|---|---|
| Qatar-4 b | 2017-02-02 | 2457787.6189 ± 0.0024 | 0.0339 ± 0.0037 |
| | 2019-05-25 | 2458628.922 ± 0.0013 | 0.0338 ± 0.0065 |
| | 2020-10-08 | 2459130.8131 ± 0.0032 | 0.0410 ± 0.0110 |
| | 2020-10-25 | 2459148.8591 ± 0.0039 | 0.0274 ± 0.0048 |
| | 2020-11-04 | 2459157.89920 ± 0.0038 | 0.0338 ± 0.0076 |
| | 2020-11-15 | 2459168.7229 ± 0.0035 | 0.0280 ± 0.0058 |
| | 2020-11-24 | 2459177.7586 ± 0.0050 | 0.0340 ± .010 |
| | 2020-12-23 | 2459206.6441 ± 0.0038 | 0.0337 ± 0.0066 |
| | 2021-01-10 | 2459224.6916 ± 0.0033 | 0.0243 ± 0.0059 |
| | 2021-10-05 | 2459493.7006 ± 0.0031 | 0.0338 ± 0.0042 |
| | 2021-11-12 | 2459531.6115 ± 0.0045 | 0.0337 ± 0.0041 |
| | 2021-12-01 | 2459549.6673 ± 0.0033 | 0.0249 ± 0.0052 |
| | 2021-12-28 | 2459576.7434 ± 0.0043 | 0.0338 ± 0.0033 |
| | 2022-02-03 | 2459614.636 ± 0.0023 | 0.0337 ± 0.0042 |
| | 2022-11-27 | 2459910.7301 ± 0.0043 | 0.0185 ± 0.0034 |
| | 2023-08-12 | 2460168.914 ± 0.011 | 0.0340 ± 0.019 |
| | 2023-11-04 | 2460253.729 ± 0.0046 | 0.0214 ± 0.0064 |
| HAT-P-18 b | 2014-02-09 | 2456708.92892 ± 0.00028 | 0.0182 ± 0.0028 |
| | 2015-09-15 | 2457281.76205 ± 0.00071 | 0.019 ± 0.0027 |
| | 2017-06-16 | 2457920.6976 ± 0.0024 | 0.0187 ± 0.0041 |
| | 2018-06-19 | 2458289.728 ± 0.0011 | 0.0224 ± 0.0018 |
| | 2019-06-23 | 2458658.7762 ± 0.0026 | 0.0204 ± 0.0061 |
| | 2020-07-07 | 2459038.8276 ± 0.0037 | 0.0121 ± 0.0031 |
| | 2020-07-30 | 2459060.8626 ± 0.0031 | 0.0219 ± 0.0046 |
| | 2021-05-06 | 2459341.767 ± 0.0038 | 0.0220 ± 0.0030 |
| | 2021-06-08 | 2459374.8207 ± 0.0026 | 0.0259 ± 0.0051 |
| | 2022-07-05 | 2459765.8753 ± 0.0030 | 0.0232 ± 0.0052 |
| CoRoT-1 b | 2013-01-04 | 2456297.6641 ± 0.0023 | 0.0205 ± 0.0022 |
| | 2013-03-31 | 2456383.676 ± 0.0021 | 0.0204 ± 0.0026 |
| | 2013-04-03 | 2456386.6926 ± 0.0024 | 0.0205 ± 0.0040 |
| | 2014-12-20 | 2457012.9137 ± 0.0029 | 0.0205 ± 0.0022 |
| | 2017-01-08 | 2457762.8687 ± 0.0037 | 0.0205 ± 0.0031 |
| | 2017-02-24 | 2457809.6417 ± 0.0033 | 0.0205 ± 0.0052 |
| | 2019-03-28 | 2458571.6797 ± 0.0045 | 0.0202 ± 0.0043 |
| | 2021-02-16 | 2459262.7837 ± 0.0052 | 0.0205 ± 0.0041 |
| | 2022-03-26 | 2459665.6746 ± 0.0055 | 0.0205 ± 0.0044 |
| | 2024-11-21* | 2460635.9566 ± 0.0024 | 0.0205 ± 0.00095 |
| | 2025-01-13* | 2460688.7483 ± 0.0015 | 0.0205 ± 0.00037 |
| | 2025-01-16* | 2460691.7671 ± 0.0018 | 0.0139 ± 0.0015 |
| | 2025-01-19* | 2460694.7945 ± 0.0012 | 0.0161 ± 0.0017 |
| | 2025-01-22* | 2460697.8098 ± 0.0012 | 0.0174 ± 0.0012 |

*Data collected from the Planewave CDK24 telescope.*

from Hartman *et al.* (2011), Kirk *et al.* (2017), Fu *et al.* (2022), Patel and Espinoza (2022), Kokori *et al.* (2022, 2023), Ivshina and Winn (2022), and Ivshina and Winn (2022) fits of Seeliger *et al.* (2015). Finally, for CoRoT-1 b, we included mid-transit times from Barge *et al.* (2008), Pont *et al.* (2010), Ivshina and Winn (2022), and Ivshina and Winn (2022) fits of Bean (2009), Gillon *et al.* (2009), Sada *et al.* (2012), and Turner *et al.* (2016). For consistency, we included only the studies that derived their published mid-transit times using measured light curves. All the previously published mid-transit times used in the creation of the O–C plots for this study can be found in Appendix C.

We used a Python code provided by Exoplanet Watch titled the "Exoplanet Ephemeris Fitting Tutorial," to generate O–C plots for the three targets and to calculate updated ephemerides. The code uses nested sampling to fit a linear ephemeris model to the observed transit times, yielding updated estimates of the orbital period and mid-transit time (Pearson

Table 2. Derived parameters from the Transiting Exoplanet Survey Satellite (TESS) transit data individual fits for Qatar-4 b, HAT-P-18 b and CoRoT-1 b.

| *Target* | *Date* | *Mid-Transit Time* ($BJD_{TDB}$) | $R_p^2/R_s^2$ | *Sector* |
|---|---|---|---|---|
| Qatar-4 b | 2024-10-02 | 2460585.9333 ± 0.0010 | 0.01689 ± 0.00095 | 84 |
| | 2024-10-04 | 2460587.7411 ± 0.0010 | 0.01573 ± 0.00091 | 84 |
| | 2024-10-06 | 2460589.54486 ± 0.00089 | 0.01703 ± 0.00088 | 84 |
| | 2024-10-07 | 2460591.3509 ± 0.0010 | 0.01582 ± 0.00083 | 84 |
| | 2024-10-09 | 2460593.15577 ± 0.00098 | 0.01759 ± 0.00087 | 84 |
| | 2024-10-11 | 2460594.96256 ± 0.00088 | 0.01704 ± 0.00084 | 84 |
| | 2024-10-15 | 2460598.57367 ± 0.00087 | 0.01885 ± 0.00090 | 84 |
| | 2024-10-16 | 2460600.37972 ± 0.0010 | 0.0165 ± 0.00087 | 84 |
| | 2024-10-18 | 2460602.18287 ± 0.00095 | 0.01605 ± 0.00089 | 84 |
| | 2024-10-22 | 2460605.79407 ± 0.00095 | 0.01924 ± 0.00091 | 84 |
| | 2024-10-24 | 2460607.59953 ± 0.00097 | 0.01688 ± 0.00093 | 84 |
| | 2024-10-25 | 2460609.40469 ± 0.00088 | 0.01717 ± 0.00088 | 84 |
| HAT-P-18 b | 2024-05-23 | 2460454.3893 ± 0.00059 | 0.0183 ± 0.00048 | 79 |
| | 2024-06-03 | 2460465.4050 ± 0.00062 | 0.01841 ± 0.00046 | 79 |
| | 2024-06-09 | 2460470.91268 ± 0.00055 | 0.01853 ± 0.00045 | 79 |
| CoRoT-1 b | 2024-12-23 | 2460667.6339 ± 0.0036 | 0.00940 ± 0.0015 | 87 |
| | 2024-12-24 | 2460669.1430 ± 0.0040 | 0.0098 ± 0.0017 | 87 |
| | 2024-12-26 | 2460670.6531 ± 0.0037 | 0.0091 ± 0.0015 | 87 |
| | 2024-12-27 | 2460672.1637 ± 0.0039 | 0.0089 ± 0.0015 | 87 |
| | 2024-12-29 | 2460673.6695 ± 0.0046 | 0.0094 ± 0.0014 | 87 |
| | 2024-12-30 | 2460675.1828 ± 0.0038 | 0.0086 ± 0.0015 | 87 |
| | 2025-01-05 | 2460681.2137 ± 0.0055 | 0.0083 ± 0.0015 | 87 |
| | 2025-01-07 | 2460682.7252 ± 0.0045 | 0.0081 ± 0.0016 | 87 |
| | 2025-01-08 | 2460684.2342 ± 0.0039 | 0.0085 ± 0.0015 | 87 |
| | 2025-01-10 | 2460685.7430 ± 0.0041 | 0.0081 ± 0.0015 | 87 |
| | 2025-01-11 | 2460687.2498 ± 0.0045 | 0.0092 ± 0.0016 | 87 |
| | 2025-01-13 | 2460688.7568 ± 0.0056 | 0.0073 ± 0.0016 | 87 |

Table 3. The updated mid-transit times and periods for Qatar-4 b, HAT-P-18 b, and CoRoT-1 b.

| *Target* | *Mid-Transit Time* ($BJD_{TDB}$) | *Period* (*days*) |
|---|---|---|
| Qatar-4 b | 2458919.5838 ± 0.000089 | 1.80536560 ± 0.00000021 |
| HAT-P-18 b | 2459743.85340 ± 0.000022 | 5.50802957 ± 0.00000012 |
| CoRoT-1 b | 2456268.99083 ± 0.000099 | 1.50896846 ± 0.000000071 |

*et al.* 2022). For each target, the most recently published values of the mid-transit time and orbital period were used as priors in the ephemeris fitting. For Qatar-4 b, we adopted the priors of $T_0$ = 2458919.58363 ± 0.00015 $BJD_{TDB}$ and P = 1.80536403 ± 0.00000054 days (Kokori *et al.* 2023). For HAT-P-18 b, the priors were $T_0$ = 2459743.853395 ± 0.000023 $BJD_{TDB}$ (Fu *et al.* 2022) and P = 5.5080232 ± 0.0000056 days (Patel and Espinoza 2022). For CoRoT-1 b, we used priors of $T_0$ = 2456268.99119 ± 0.00012 $BJD_{TDB}$ and P = 1.508968772 ± 0.000000083 days (Ivshina and Winn 2022). The O–C plots and posterior distributions are presented in Figures 4 and 5, respectively.

## 5. Results

In this study, we present updated mid-transit times and periods for three planets: Qatar-4 b, HAT-P-18 b, and CoRoT-1 b, derived by integrating data collected from small, ground-based, robotic telescopes, new TESS data, and archival published results. The results are summarized in Table 3.

We compared our updated mid-transit time uncertainties and orbital period uncertainties to those most recently published in prior studies. Our measurements reduce Qatar-4 b's mid-transit time uncertainty by 36.4% and its orbital period uncertainty by 65.0% relative to the most recently published values (Kokori *et al.* 2023). HAT-P-18 b's mid-transit time uncertainty was reduced slightly by 4.35% when compared to Fu *et al.* (2022), a study that utilized an observation collected using JWST Near Infrared Imager and Slitless Spectrograph single object slitless spectroscopy (NIRISS/SOSS) to characterize the planet's atmosphere and present an updated ephemeris. The orbital period uncertainty was reduced by 77.4% compared to the most recently published value (Kokori *et al.* 2023). Finally, CoRoT-1 b's mid-transit uncertainty was reduced by 17.5% and its orbital period uncertainty by 16.9% relative to those presented in Ivshina and Winn (2022).

In addition to the direct comparisons, we also forward-propagated the mid-transit time uncertainties reported in this study along with previously published mid-transit time uncertainties to a potential start date of the *Ariel* science missions (adopting 2031 January 1), in order to investigate the potential impact our study has on future observations of Qatar-4 b, HAT-P-18 b, and CoRoT-1 b. We calculated the planets' mid-transit time uncertainties for the future date using Equation 1 which is analytically derived in Zellem *et al.* (2020):

$$\Delta T_{mid} = (n_{orbit}^2 \cdot \Delta P^2 + 2n_{orbit} \cdot \Delta P \Delta T_0 + T_0^2)^{1/2} \qquad (1)$$

where $n_{orbit}$ is the number of orbits that will occur before the future date, P is the orbital period of the planet, and $T_0$ is the

Table 4. Mid-transit times and orbital periods from this work and the most recent previous studies, along with the resulting mid-transit time uncertainties forward propagated to an estimated Ariel mission start date of 2031 January 1 (2462867.50 BJDTDB).

| *Target* | *Source* | *Mid-Transit Time* ($BJD_{TDB}$) | *Period* (*days*) | *Propagated Mid-Transit Time Uncertainty* ($BJD_{TDB}$) |
|---|---|---|---|---|
| Qatar-4 b | This Work | 2458919.5838 ± 0.000089 | 1.80536560 ± 0.00000021 | 0.00047 |
| | Kokori *et al.* (2023) | 2458919.58363 ± 0.00015 | 1.80536403 ± 0.00000054 | 0.0012 |
| HAT-P-18 b | This Work | 2459743.85340 ± 0.000022 | 5.50802957 ± 0.00000012 | 0.000072 |
| | Fu *et al.* (2022)* | 2459743.853395 ± 0.000023 | 5.50802941 ± 0.00000053 | 0.00030 |
| CoRoT-1 b | This Work | 2456268.99083 ± 0.000099 | 1.50896846 ± 0.000000071 | 0.00033 |
| | Ivshina and Winn (2022) | 2456268.99119 ± 0.00012 | 1.508968772 ± 0.000000083 | 0.00038 |

* *For HAT-P-18 b, the orbital period was adopted from Kokori et al. (2023), as Fu et al. (2022) did not report a period.*

established mid-transit time. The second term in Equation 1 was dropped since none of the previous studies reported their covariance term. Dropping the covariance term will lead to a slight underestimation of the resulting uncertainty in the mid-transit time. However, because our updated period uncertainties are substantially smaller than previous values, the effect on comparative forward-propagated uncertainties is minimal. We applied this forward propagation both to the mid-transit time uncertainties derived in this study and to the most recently published ephemerides for each target. The values used in the calculations as well as the final results are presented in Table 4.

The results of the forward propagation demonstrate how the updated mid-transit times presented in this study yield lower uncertainties in the propagated mid-transit times with the incorporation of more data. Our results for Qatar-4 b yield a 61% smaller uncertainty in the propagated mid-transit time. Similar improvements are found for HAT-P-18 b and CoRoT-1 b, with reductions of 76% and 13%, respectively.

## 6. Conclusion

In this study, we presented updated mid-transit times and orbital periods for three hot Jupiters: Qatar-4 b, HAT-P-18 b, and CoRoT-1 b, by integrating data collected from small, ground-based observatories and personal telescopes, unreduced TESS data, and previously published results. We reported an updated mid-transit time of 2458919.5838 ± 0.000089 $BJD_{TDB}$ and orbital period of 1.80536560 ± 0.00000021 days for Qatar-4b, a mid-transit time of 2459743.85340 ± 0.000022 $BJD_{TDB}$ and orbital period of 5.50802957 ± 0.00000012 days for HAT-P-18 b, and a midtransit time of 2456268.99083 ± 0.000099 $BJD_{TDB}$ and orbital period of 1.50896846 ± 0.000000071 days for CoRoT-1 b. For all three targets, the updated ephemerides demonstrate improved precision compared to previous published values. These improvements demonstrate the continued value of long-term, ground-based monitoring and the importance of integrating multiple data sources in maintaining accurate exoplanet ephemerides.

To assess the long-term impact of these updated ephemerides, we forward-propagated the mid-transit time uncertainties found in this study to a potential start date of the *Ariel* science operations (2031 January 1). The updated ephemerides presented in this study yield smaller propagated mid-transit time uncertainties compared to those derived from previously published ephemerides, underscoring the importance of regular ephemeris refinement. These improvements are particularly relevant for hot Jupiters, whose short orbital periods cause uncertainties to accumulate rapidly and can otherwise compromise the efficiency of time-critical observations with space-based facilities.

The results of this study highlight the continued value of coordinated, long-term ground-based monitoring programs and demonstrate that ephemeris refinement can be achieved by combining ground-based telescope data, space-based survey data, and previously published results. Citizen science initiatives, such as Exoplanet Watch, play a critical role in maintaining accurate exoplanet ephemerides, helping to minimize the temporal and financial constraints of using large ground-based and space-based observatories.

## 7. Acknowledgements

The observations in this work were conducted with MicroObservatory, a robotic telescope network maintained and operated as an educational service by the Center for Astrophysics|Harvard and Smithsonian. MicroObservatory is supported by NASA's Universe of Learning under NASA award number NNX16AC65A to the Space Telescope Science Institute.

This publication makes use of data products from Exoplanet Watch, a citizen science project managed by NASA's Jet Propulsion Laboratory on behalf of NASA's Universe of Learning. This work is supported by NASA under award number NNX16AC65A to the Space Telescope Science Institute, in partnership with Caltech / IPAC, Center for Astrophysics| Harvard and Smithsonian, and NASA Jet Propulsion Laboratory. This research has made use of NASA's Astrophysics Data System. Some/all of the data presented in this paper were obtained from the Mikulski Archive for Space Telescopes (MAST). STScI is operated by the Association of Universities for Research in Astronomy, Inc., under NASA contract NAS5-26555. Support for MAST for non-HST data is provided by the NASA Office of Space Science via grant NNX13AC07G and by other grants and contracts.

This research has made use of the NASAExoplanetArchive, which is operated by the California Institute of Technology, under contract with the National Aeronautics and Space Administration under the Exoplanet Exploration Program.

This work is supported by the National Science Foundation (NSF) under grant #IUSE 2121225.

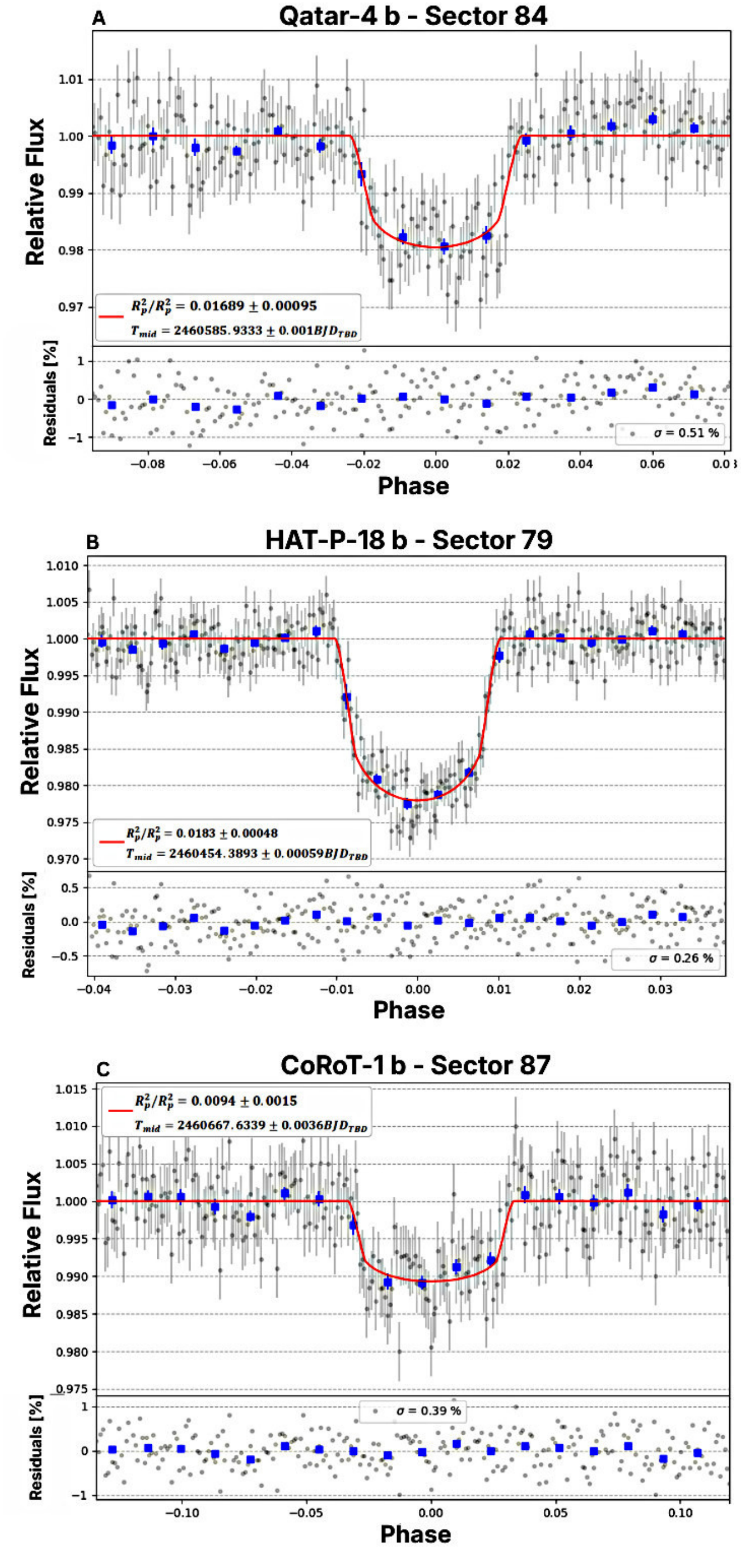

Figure 3. Example light curves reduced using TESS data for each target. Qatar-4 b (A) was observed on 2024-10-02, HAT-P-18 b (B) was observed on 2024-05-23, and CoRoT-1 b (C) was observed on 2024-12-23.

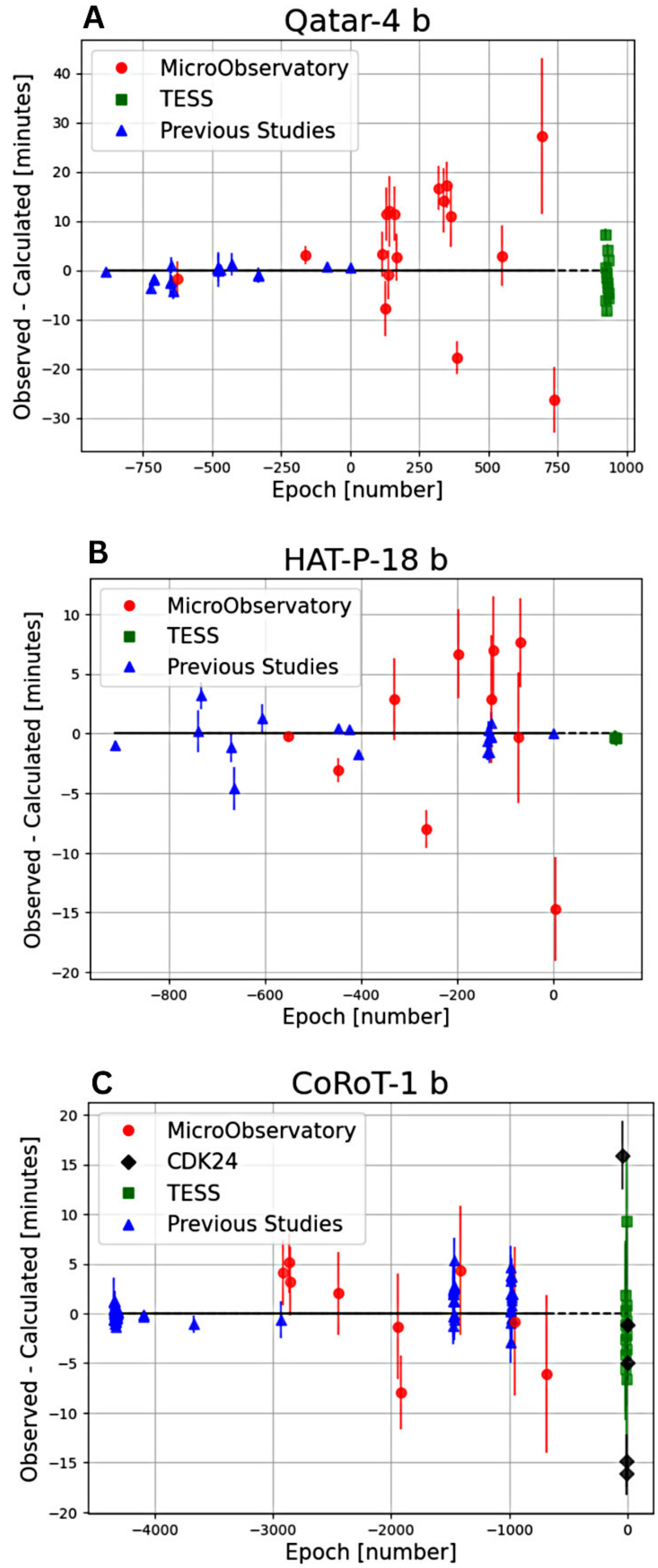

Figure 4. O–C plots for Qatar-4 b (A), HAT-P-18 b (B), and CoRoT-1 b (C) using ground-based and TESS data as well as previously published values. The priors used for (A) were $T_0 = 2458919.58363 \pm 0.00015$ $BJD_{TDB}$ and $P = 1.80536403 \pm 0.00000054$ days, the priors used for (B) were $T_0 = 2459743.853395 \pm 0.000023$ $BJD_{TDB}$ and $P = 5.5080232 \pm 0.0000056$ days, and the priors used for (C) $T_0 = 2456268.99119 \pm 0.00012$ $BJD_{TDB}$ and $P = 1.508968772 \pm 0.000000083$ days.

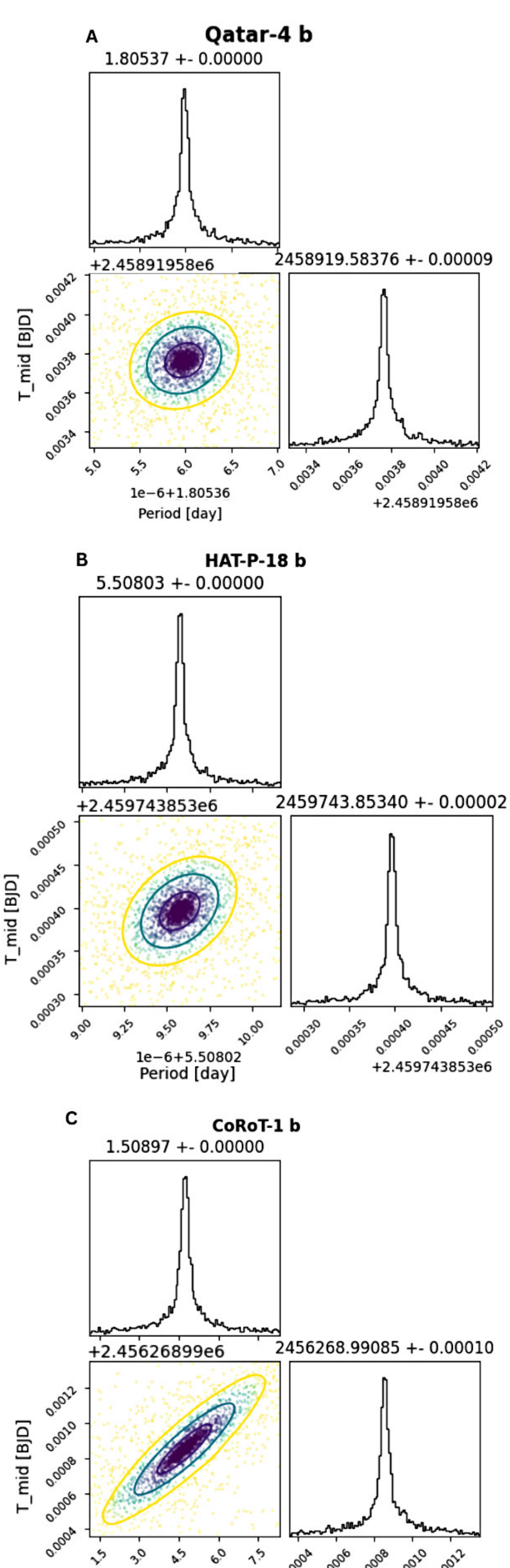

Figure 5. The posterior plot distributions for our newly calculated mid-transit times and orbital periods for Qatar-4 b (A), HAT-P-18 b (B), and CoRoT-1 b (C).

[1] NASA (2026): https://science.nasa.gov/citizen-science/exoplanet-watch/exoplanet-watch-overview

**Appendix A: Significant detections of Qatar-4 b, HAT-P-18 b, and CoRoT-1 b, reduced using ground-based data**

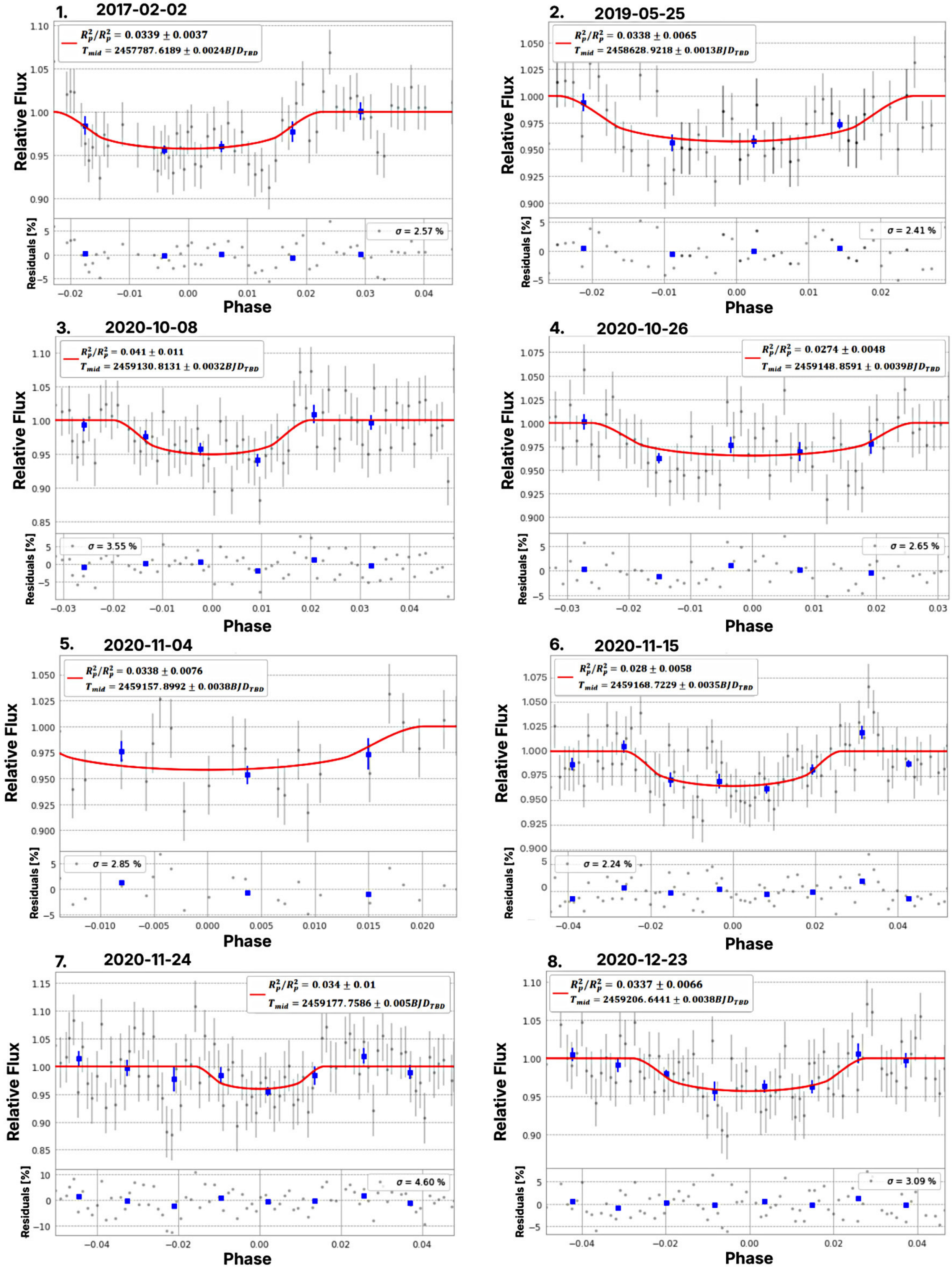

*Continued on following pages*

Appendix A: Significant detections of Qatar-4 b, HAT-P-18 b, and CoRoT-1 b, reduced using ground-based data, cont.

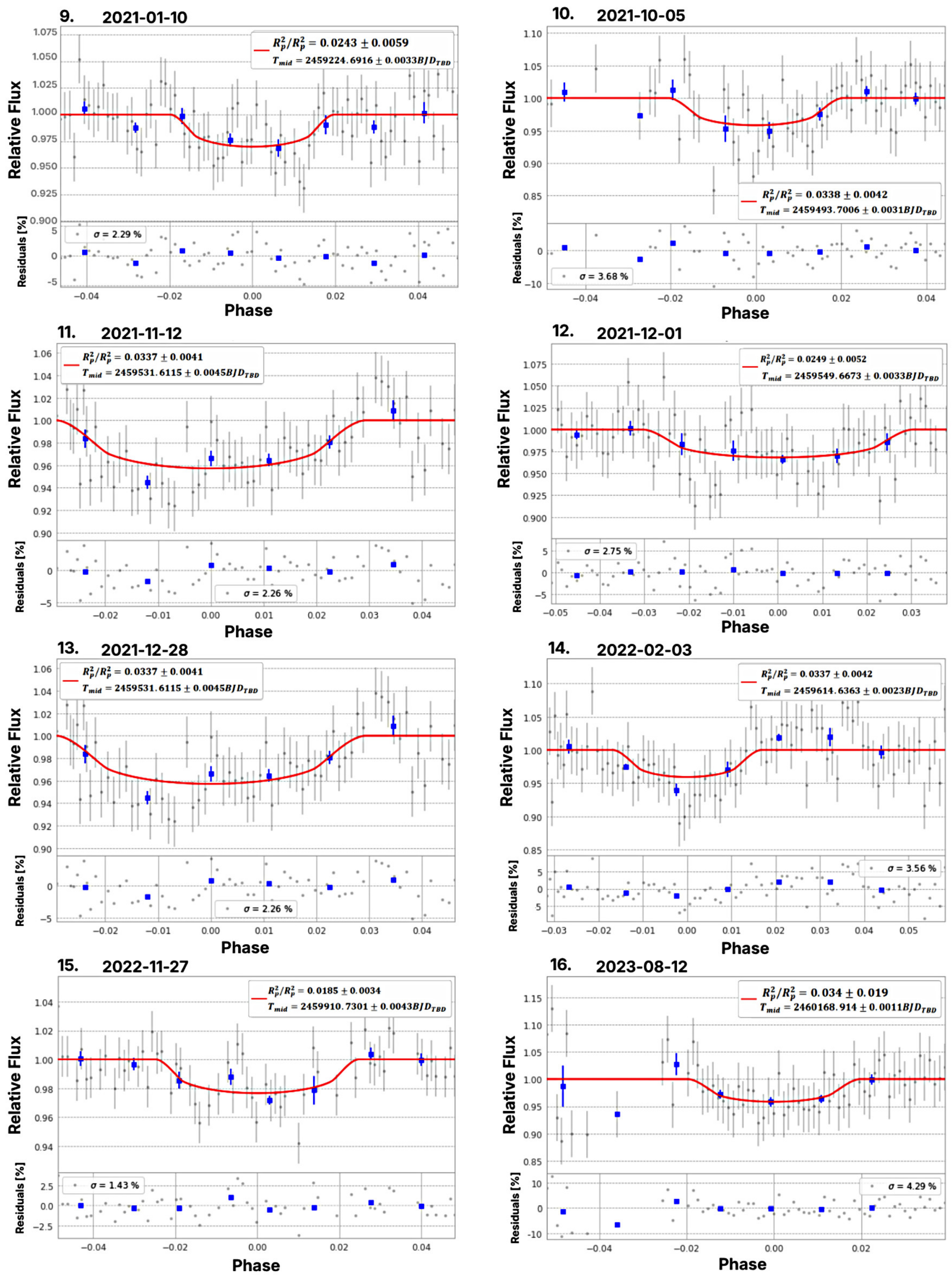

*Continued on following pages*

Appendix A: Significant detections of Qatar-4 b, HAT-P-18 b, and CoRoT-1 b, reduced using ground-based data, cont.

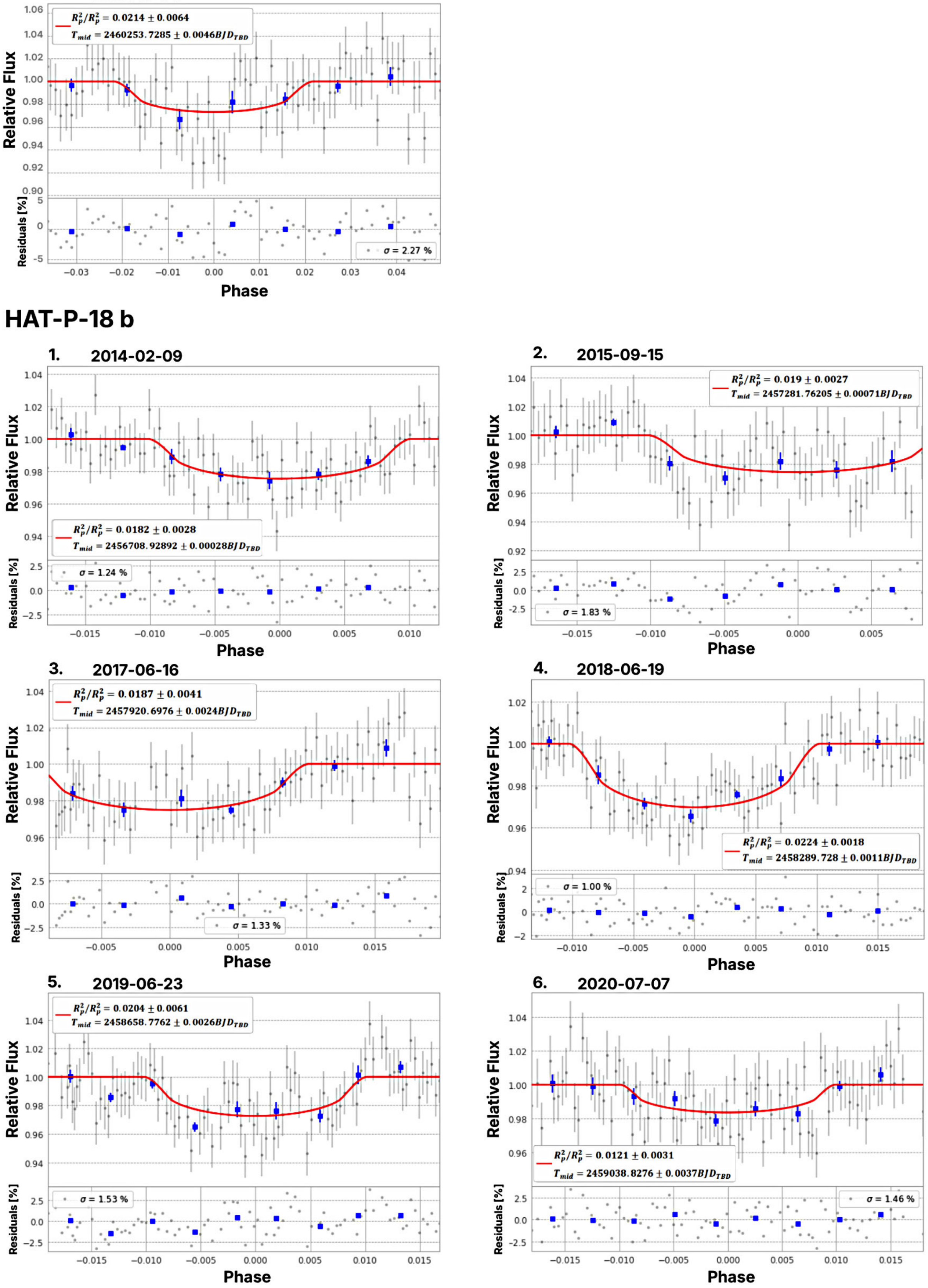

*Continued on following pages*

Appendix A: Significant detections of Qatar-4 b, HAT-P-18 b, and CoRoT-1 b, reduced using ground-based data, cont.

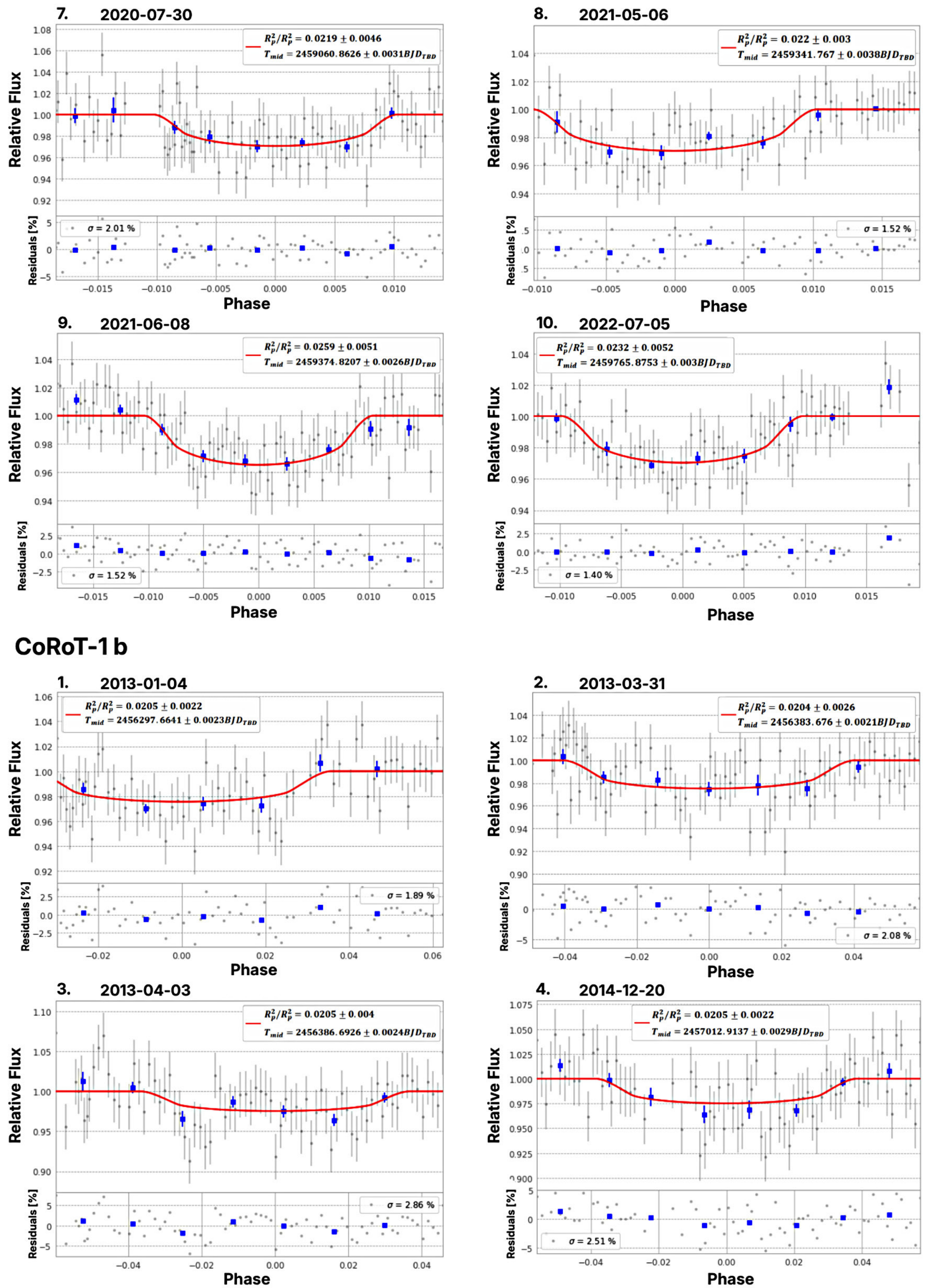

Continued on following pages

Appendix A: Significant detections of Qatar-4 b, HAT-P-18 b, and CoRoT-1 b, reduced using ground-based data, cont.

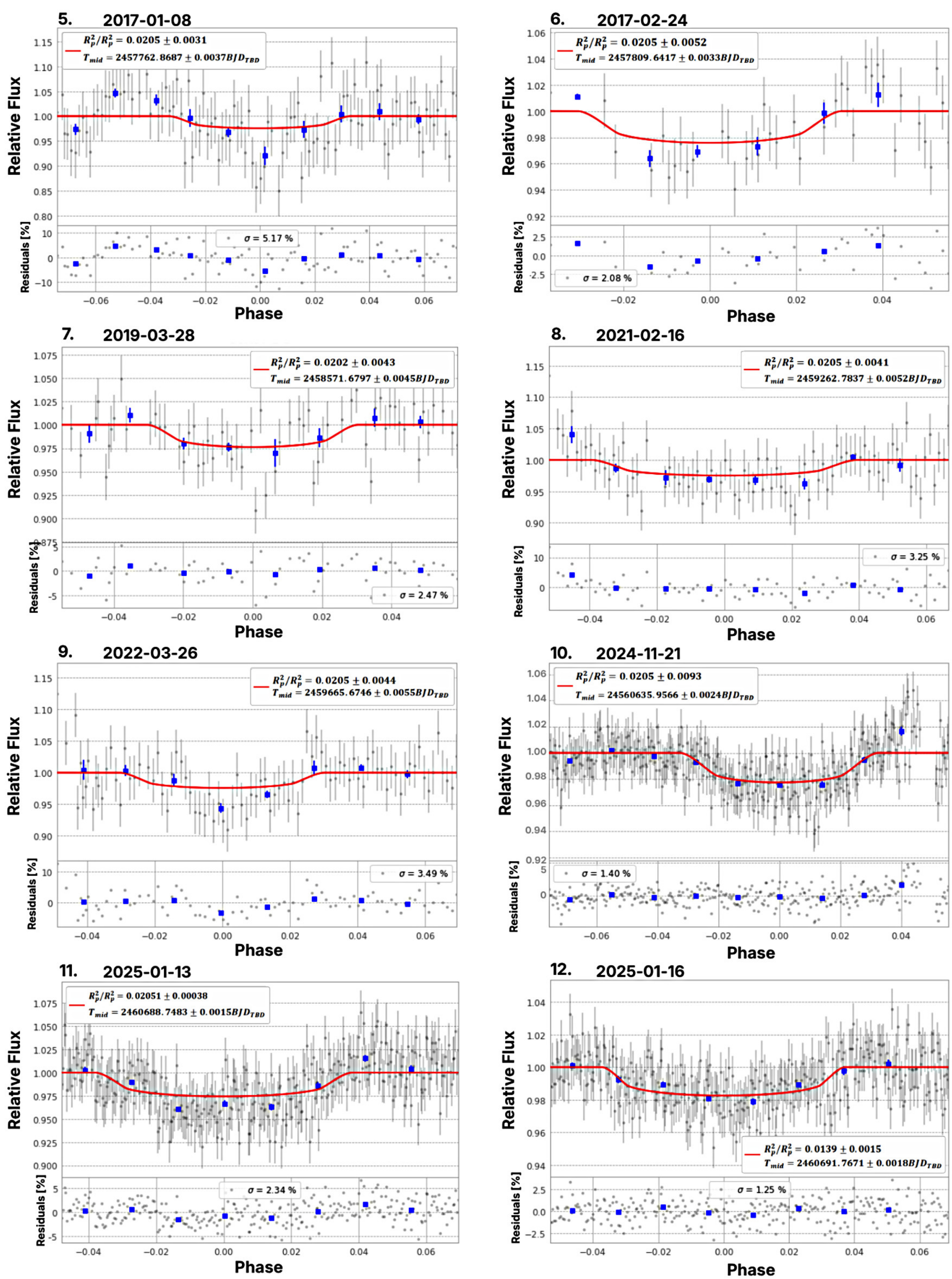

*Continued on next page*

Appendix A: Significant detections of Qatar-4 b, HAT-P-18 b, and CoRoT-1 b, reduced using ground-based data, cont.

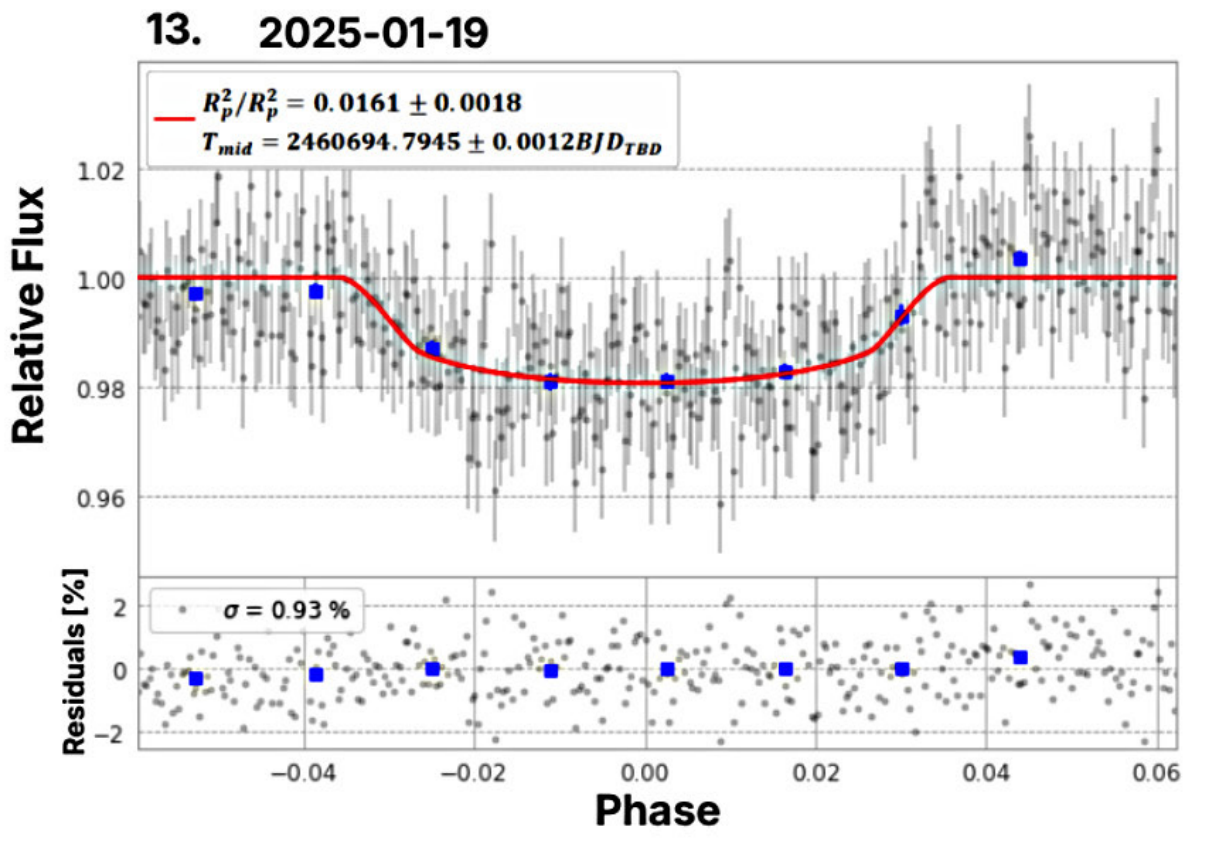

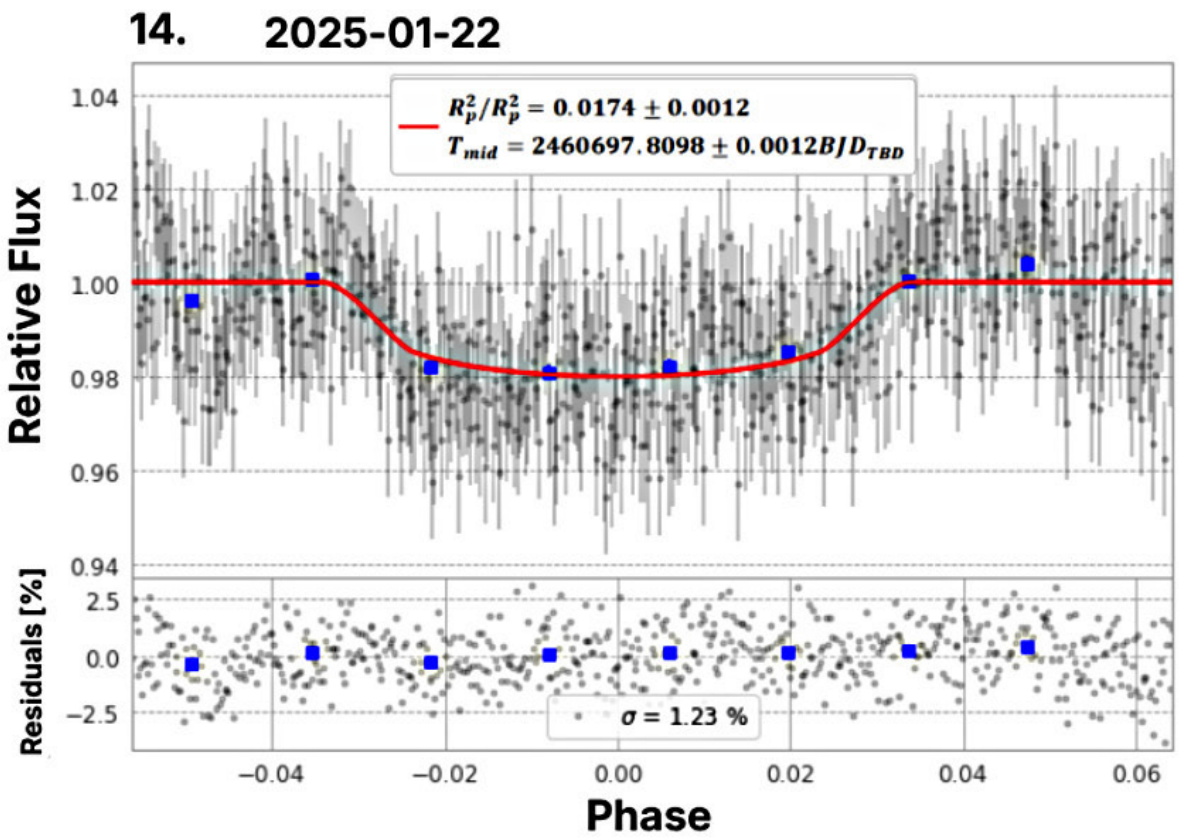

**Appendix B: Significant detections of Qatar-4 b, HAT-P-18 b, and CoRoT-1 b, reduced using TESS data**

## Qatar-4 b

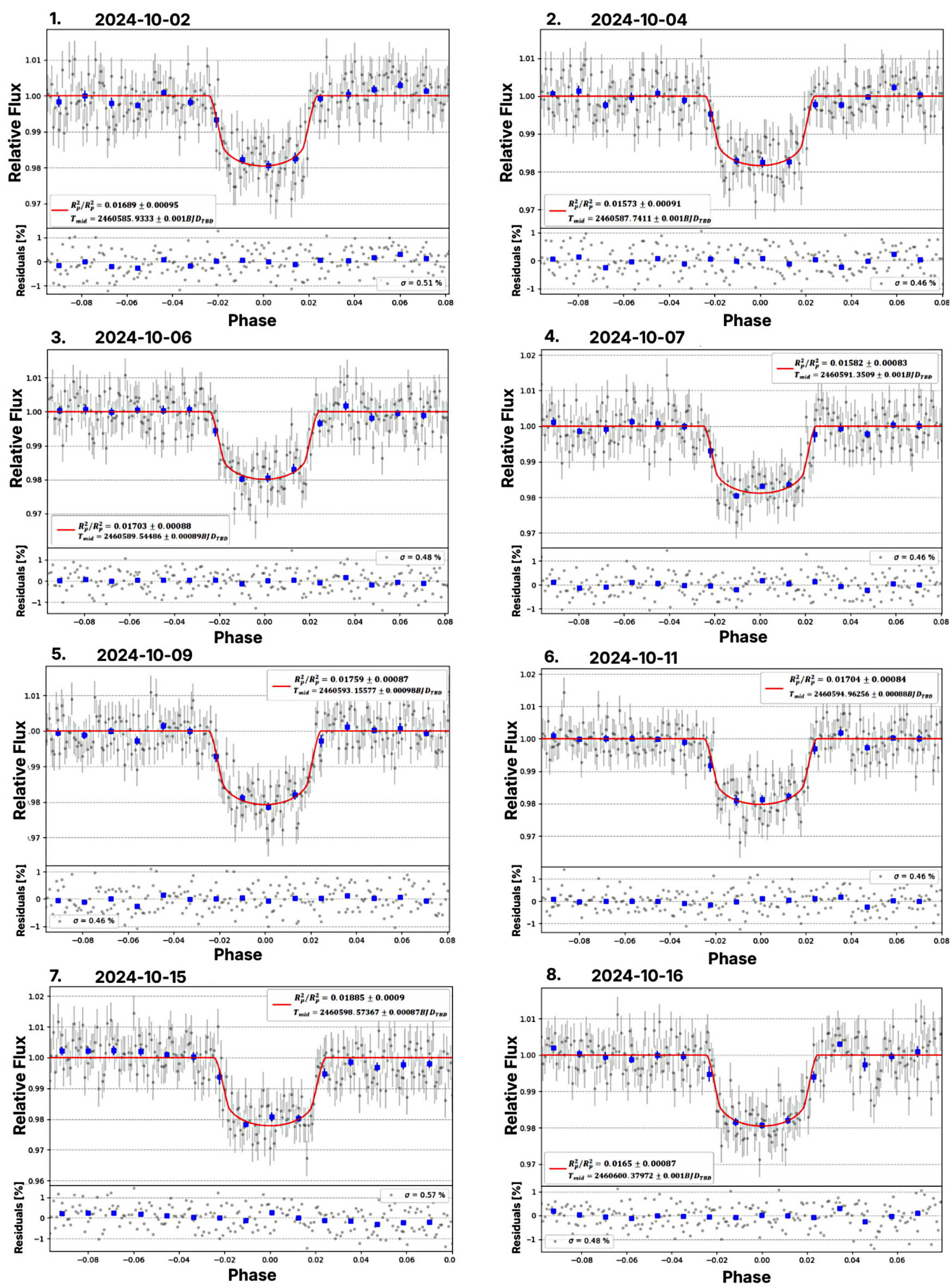

*Continued on following pages*

Appendix B: Significant detections of Qatar-4 b, HAT-P-18 b, and CoRoT-1 b, reduced using TESS data, cont.

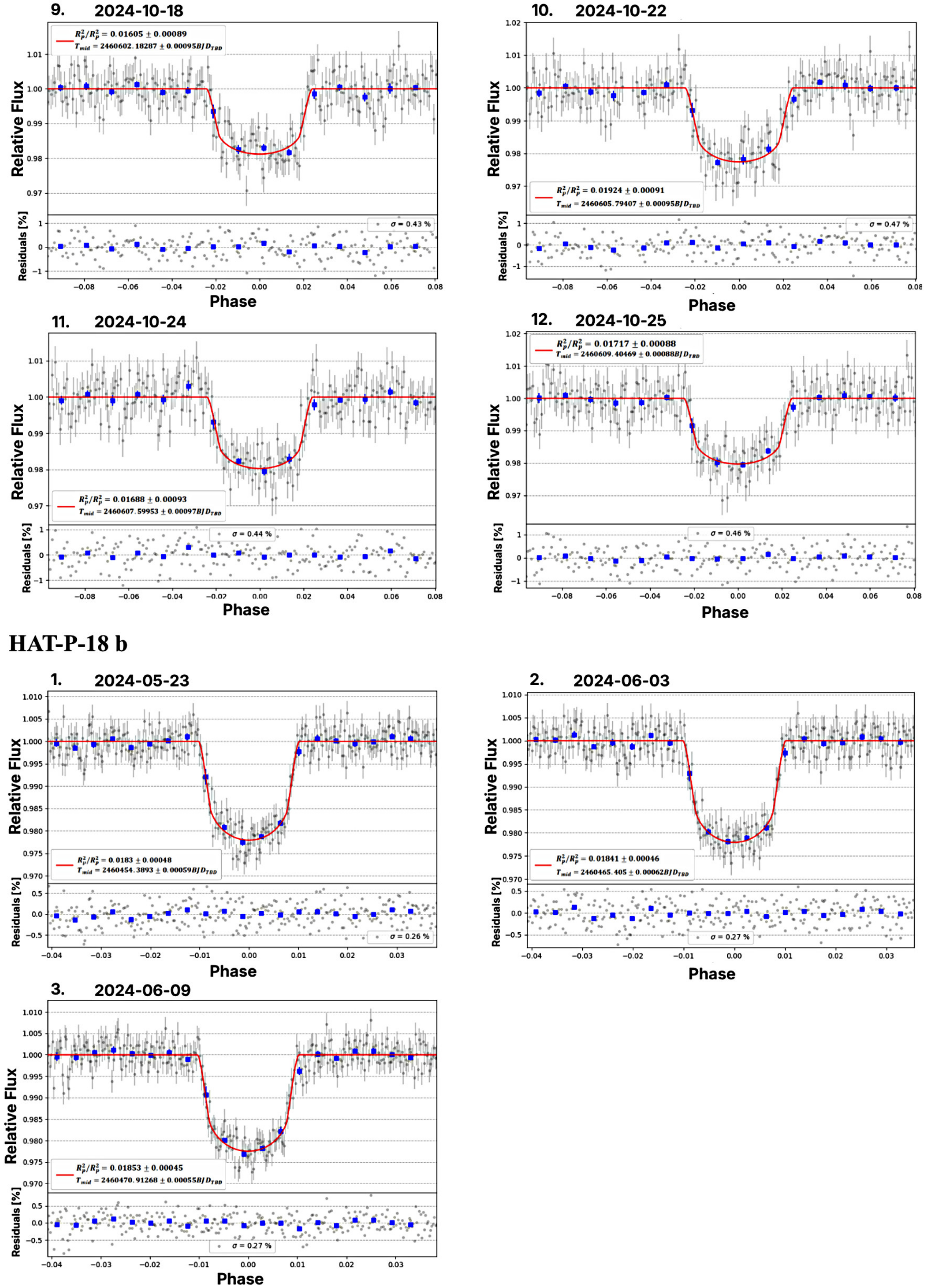

*Continued on following pages*

Appendix B: Significant detections of Qatar-4 b, HAT-P-18 b, and CoRoT-1 b, reduced using TESS data, cont.

## CoRoT-1 b

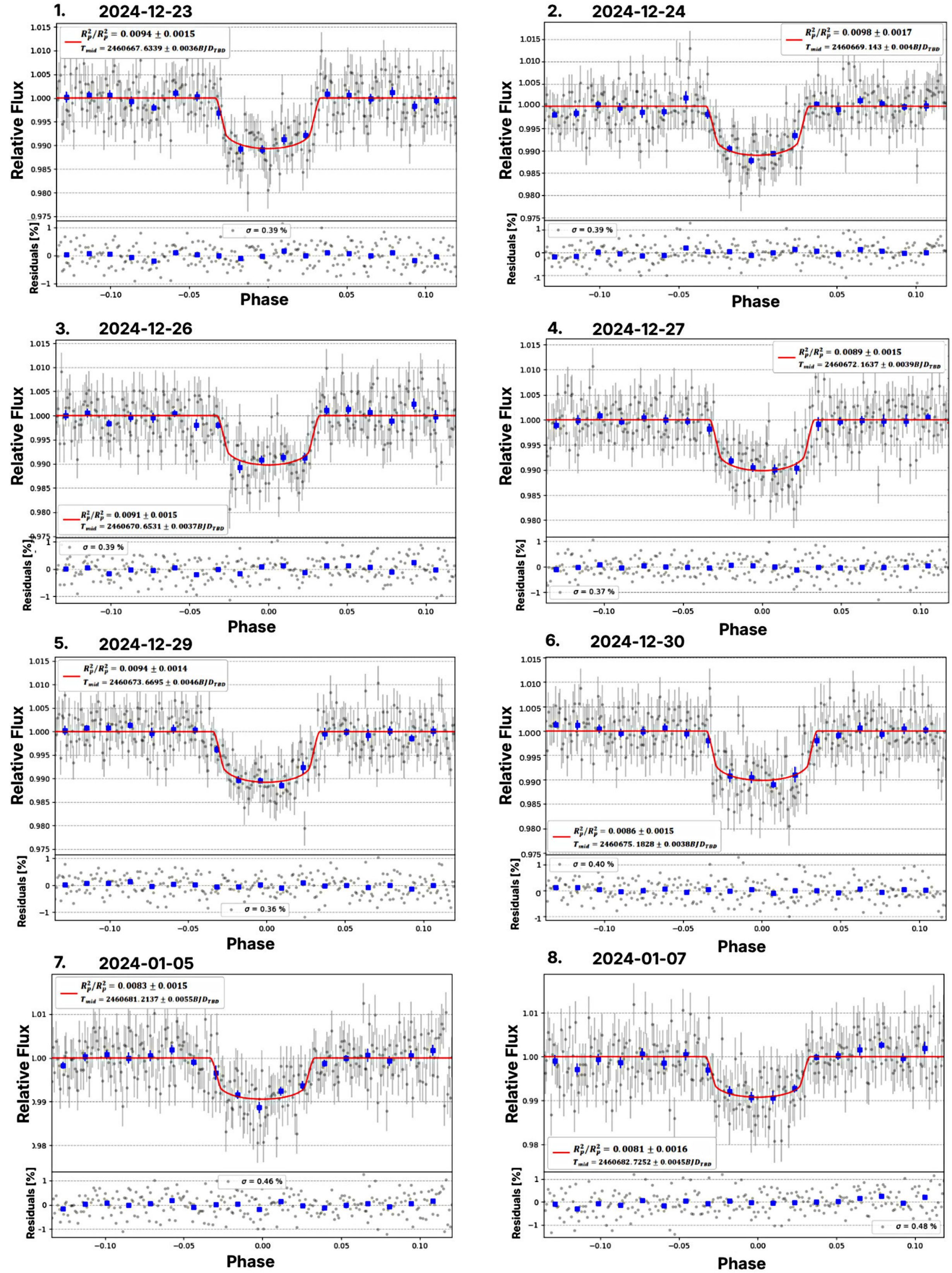

*Continued on next page*

Appendix B: Significant detections of Qatar-4 b, HAT-P-18 b, and CoRoT-1 b, reduced using TESS data, cont.

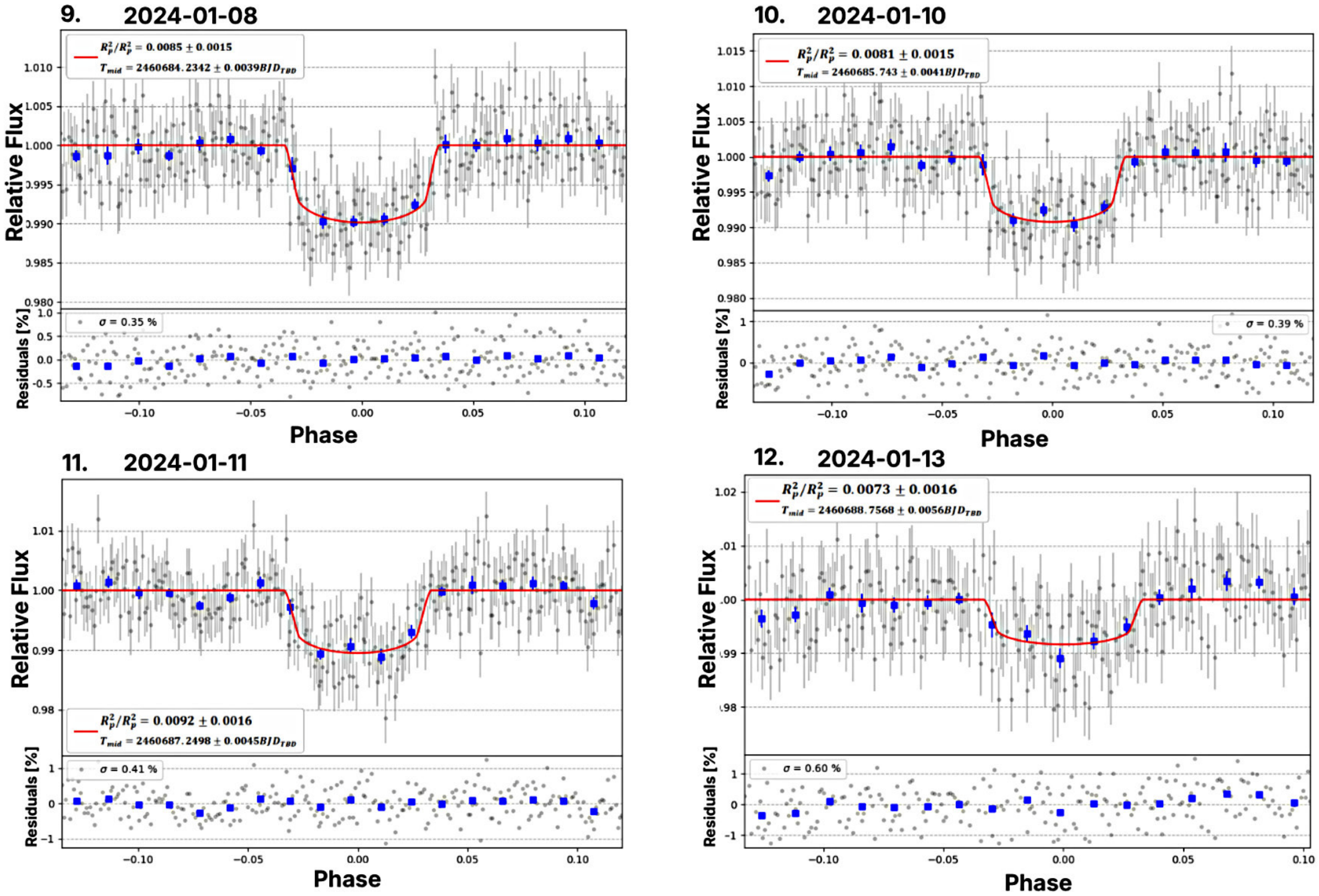

## Appendix C: Prior solutions included in the O–C plots for Qatar-4 b, HAT-P-18 b, and CoRoT-1 b (Figure 4)

Table C1. Prior solutions included in the O–C plots for Qatar-4 b, HAT-P-18 b, and CoRoT-1 b (Figure 4).

| *Target* | *Mid-Transit Time ($BJD_{TDB}$)* | *Reference* |
|---|---|---|
| Qatar-4 b | 2457323.64130 ± 0.0004895 | Ivshina and Winn (2022) fit of Wang *et al.* (2021) |
| | 2457617.91334 ± 0.0005241 | Ivshina and Winn (2022) fit of Wang *et al.* (2021) |
| | 2457637.77349 ± 0.0005464 | Ivshina and Winn (2022) fit of Wang *et al.* (2021) |
| | 2457637.77361 ± 0.00046 | Ivshina and Winn (2022) fit of Mallonn *et al.* (2019) and Alsubai *et al.* (2017) |
| | 2457744.28950 ± 0.00164 | Ivshina and Winn (2022) fit of Mallonn *et al.* (2019) |
| | 2457753.31672 ± 0.00074 | Ivshina and Winn (2022) fit of Mallonn *et al.* (2019) |
| | 2457753.31884 ± 0.00115 | Ivshina and Winn (2022) fit of Mallonn *et al.* (2019) |
| | 2457762.34209 ± 0.00125 | Ivshina and Winn (2022) fit of Mallonn *et al.* (2019) |
| | 2458049.39801 ± 0.00058 | Ivshina and Winn (2022) fit of Mallonn *et al.* (2019) |
| | 2458058.42459 ± 0.00235 | Ivshina and Winn (2022) fit of Mallonn *et al.* (2019) |
| | 2458058.42511 ± 0.00203 | Ivshina and Winn (2022) fit of Mallonn *et al.* (2019) |
| | 2458071.06228 ± 0.0007135 | Ivshina and Winn (2022) fit of Wang *et al.* (2021) |
| | 2458143.27752 ± 0.00082 | Ivshina and Winn (2022) fit of Mallonn *et al.* (2019) |
| | 2458143.27761 ± 0.00159 | Ivshina and Winn (2022) fit of Mallonn *et al.* (2019) |
| | 2458316.59098 ± 0.000516 | Ivshina and Winn (2022) fit of Wang *et al.* (2021) |
| | 2458316.59101 ± 0.00103 | Ivshina and Winn (2022) fit of Mallonn *et al.* (2019) |
| | 2458767.93317 ± 0.00014 | Kokori *et al.* (2022) |
| | 2458919.58363 ± 0.00015 | Kokori *et al.* (2023) |
| HAT-P-18 b | 2454715.02174 ± 0.00020 | Hartman *et al.* (2011) |
| | 2457408.449133 ± 0.000081 | Kokori *et al.* (2023) |
| | 2459743.853395 ± 0.000023 | Fu *et al.* (2022) |
| | $2459033.31735^{+0.00021}_{-0.00022}$ | Patel and Espinoza (2022) |
| | 2457276.25646 ± 0.00010 | Kokori *et al.* (2022) |
| | $2457507.59219566^{+0.00019400}_{-0.00018100}$ | Kirk *et al.* (2017) |
| | 2455673.41967 ± 0.00124 | Ivshina and Winn (2022) fit of Seeliger *et al.* (2015) |
| | 2455706.46993 ± 0.0008 | Ivshina and Winn (2022) fit of Seeliger *et al.* (2015) |
| | 2456053.47276 ± 0.00084 | Ivshina and Winn (2022) fit of Seeliger *et al.* (2015) |
| | 2456086.51856 ± 0.00125 | Ivshina and Winn (2022) fit of Seeliger *et al.* (2015) |
| | 2456411.49638 ± 0.00084 | Ivshina and Winn (2022) fit of Seeliger *et al.* (2015) |
| | 2458989.25287 ± 0.0006096 | Ivshina and Winn (2022) |
| | 2458994.76030 ± 0.0004839 | Ivshina and Winn (2022) |
| | 2459005.77761 ± 0.0005205 | Ivshina and Winn (2022) |
| | 2459011.28525 ± 0.0007245 | Ivshina and Winn (2022) |
| | 2459016.79238 ± 0.0006163 | Ivshina and Winn (2022) |
| | 2459027.80945 ± 0.0006292 | Ivshina and Winn (2022) |
| | 2459033.31820 ± 0.0006397 | Ivshina and Winn (2022) |
| CoRoT-1 b | 2454138.32761 ± 0.00047 | Ivshina and Winn (2022) fit of Bean, (2009) |
| | 2454139.83712 ± 0.00059 | Ivshina and Winn (2022) fit of Bean, (2009) |
| | 2454141.34485 ± 0.00062 | Ivshina and Winn (2022) fit of Bean, (2009) |
| | 2454142.85425 ± 0.00039 | Ivshina and Winn (2022) fit of Bean, (2009) |
| | 2454144.36411 ± 0.00163 | Ivshina and Winn (2022) fit of Bean, (2009) |
| | 2454145.87255 ± 0.00042 | Ivshina and Winn (2022) fit of Bean, (2009) |
| | 2454147.38076 ± 0.00048 | Ivshina and Winn (2022) fit of Bean, (2009) |
| | 2454148.88942 ± 0.00038 | Ivshina and Winn (2022) fit of Bean, (2009) |
| | 2454150.39899 ± 0.00026 | Ivshina and Winn (2022) fit of Bean, (2009) |
| | 2454151.90798 ± 0.00045 | Ivshina and Winn (2022) fit of Bean, (2009) |
| | 2454153.41661 ± 0.00041 | Ivshina and Winn (2022) fit of Bean, (2009) |
| | 2454154.92615 ± 0.00035 | Ivshina and Winn (2022) fit of Bean, (2009) |
| | 2454156.43527 ± 0.00026 | Ivshina and Winn (2022) fit of Bean, (2009) |
| | 2454157.94481 ± 0.00073 | Ivshina and Winn (2022) fit of Bean, (2009) |
| | 2454159.4532 ± 0.0001 | Barge *et al.* (2008) |
| | 2454159.45331 ± 0.0005 | Ivshina and Winn (2022) fit of Bean, (2009) |
| | 2454160.96252 ± 0.00029 | Ivshina and Winn (2022) fit of Bean, (2009) |
| | 2454162.47066 ± 0.00045 | Ivshina and Winn (2022) fit of Bean, (2009) |
| | 2454163.97926 ± 0.00044 | Ivshina and Winn (2022) fit of Bean, (2009) |
| | 2454165.48852 ± 0.00047 | Ivshina and Winn (2022) fit of Bean, (2009) |
| | 2454166.99747 ± 0.00054 | Ivshina and Winn (2022) fit of Bean, (2009) |
| | 2454168.50619 ± 0.00028 | Ivshina and Winn (2022) fit of Bean, (2009) |
| | 2454170.01596 ± 0.00020 | Ivshina and Winn (2022) fit of Bean, (2009) |

*Table continued on next page*

Table C1. Prior solutions included in the O–C plots for Qatar-4 b, HAT-P-18 b, and CoRoT-1 b (Figure 4), cont.

| *Target* | *Mid-Transit Time* ($BJD_{TDB}$) | *Reference* |
|---|---|---|
| CoRoT-1 b | 2454171.52372 ± 0.00028 | Ivshina and Winn (2022) fit of Bean, (2009) |
| | 2454173.03365 ± 0.00026 | Ivshina and Winn (2022) fit of Bean, (2009) |
| | 2454174.54240 ± 0.00033 | Ivshina and Winn (2022) fit of Bean, (2009) |
| | 2454176.05183 ± 0.00026 | Ivshina and Winn (2022) fit of Bean, (2009) |
| | 2454177.56026 ± 0.00023 | Ivshina and Winn (2022) fit of Bean, (2009) |
| | 2454179.06932 ± 0.00029 | Ivshina and Winn (2022) fit of Bean, (2009) |
| | 2454180.57844 ± 0.00025 | Ivshina and Winn (2022) fit of Bean, (2009) |
| | 2454183.59609 ± 0.00024 | Ivshina and Winn (2022) fit of Bean, (2009) |
| | 2454185.10512 ± 0.00031 | Ivshina and Winn (2022) fit of Bean, (2009) |
| | 2454186.61471 ± 0.00031 | Ivshina and Winn (2022) fit of Bean, (2009) |
| | 2454188.12338 ± 0.0005 | Ivshina and Winn (2022) fit of Bean, (2009) |
| | 2454189.63227 ± 0.00038 | Ivshina and Winn (2022) fit of Bean, (2009) |
| | 2454191.14143 ± 0.00026 | Ivshina and Winn (2022) fit of Bean, (2009) |
| | 2454524.6231 ± 0.0002 | Pont *et al.* (2010) |
| | 2454524.62324 ± 0.00013 | Ivshina and Winn (2022) fit of Gillon *et al.* (2009) |
| | 2455162.91621 ± 0.00059 | Ivshina and Winn (2022) fit of Sada *et al.* (2012) |
| | 2456268.99040 ± 0.0013 | Ivshina and Winn (2022) fit of Turner *et al.* (2016) |
| | 2458469.06821 ± 0.0012333 | Ivshina and Winn (2022) |
| | 2458470.57717 ± 0.0013167 | Ivshina and Winn (2022) |
| | 2458472.08565 ± 0.0014233 | Ivshina and Winn (2022) |
| | 2458473.59354 ± 0.0014722 | Ivshina and Winn (2022) |
| | 2458475.10442 ± 0.0014377 | Ivshina and Winn (2022) |
| | 2458476.61084 ± 0.0013014 | Ivshina and Winn (2022) |
| | 2458479.63123 ± 0.0019923 | Ivshina and Winn (2022) |
| | 2458481.13813 ± 0.0013734 | Ivshina and Winn (2022) |
| | 2458482.64751 ± 0.0014385 | Ivshina and Winn (2022) |
| | 2458484.16026 ± 0.001535 | Ivshina and Winn (2022) |
| | 2458485.66746 ± 0.0014853 | Ivshina and Winn (2022) |
| | 2458487.17535 ± 0.001727 | Ivshina and Winn (2022) |
| | 2458488.68526 ± 0.0014447 | Ivshina and Winn (2022) |
| | 2459202.42349 ± 0.001405 | Ivshina and Winn (2022) |
| | 2459203.93381 ± 0.0013578 | Ivshina and Winn (2022) |
| | 2459205.44363 ± 0.0010803 | Ivshina and Winn (2022) |
| | 2459206.95508 ± 0.0014073 | Ivshina and Winn (2022) |
| | 2459208.46275 ± 0.0011239 | Ivshina and Winn (2022) |
| | 2459209.97053 ± 0.0014639 | Ivshina and Winn (2022) |
| | 2459211.48163 ± 0.0013722 | Ivshina and Winn (2022) |
| | 2459212.99149 ± 0.0015417 | Ivshina and Winn (2022) |
| | 2459216.00736 ± 0.0010382 | Ivshina and Winn (2022) |
| | 2459217.51579 ± 0.0014488 | Ivshina and Winn (2022) |
| | 2459219.02554 ± 0.0012323 | Ivshina and Winn (2022) |
| | 2459220.53278 ± 0.0018204 | Ivshina and Winn (2022) |
| | 2459222.04305 ± 0.0013553 | Ivshina and Winn (2022) |
| | 2459223.55080 ± 0.0016507 | Ivshina and Winn (2022) |
| | 2459225.06259 ± 0.0013206 | Ivshina and Winn (2022) |
| | 2459226.57034 ± 0.0013817 | Ivshina and Winn (2022) |